\documentclass[prb,twocolumn]{revtex4}
\usepackage{bm}
\usepackage{graphicx}
\usepackage{amssymb}
\usepackage{amsmath}
\usepackage{eufrak}
\usepackage{color}
\usepackage[utf8]{inputenc} 
\usepackage{hyperref}
\usepackage{pifont}
\usepackage{ulem}
\usepackage{array}
\usepackage{verbatim}

\begin{document}
	
\title{Photon Drag Effect in (Bi$_{1-x}$Sb$_{x}$)$_{2}$Te$_{3}$ Three Dimensional Topological Insulators}
	\author{ H.~Plank$^1$, L.\,E.~Golub$^2$, S.~Bauer$^1$, V.\,V.~Bel'kov$^2$, T.~Herrmann$^1$, P.~Olbrich$^1$, M.~Eschbach$^3$, L.~Plucinski$^3$, J.~Kampmeier$^3$, M.~Lanius$^3$, G.~Mussler$^3$, D.~Gr\"{u}tzmacher$^3$, and S.\,D.~Ganichev$^1$	}
	\affiliation{$^1$Terahertz Center, University of Regensburg, 93040 Regensburg, Germany}
	\affiliation{$^2$Ioffe 
		Institute,
		194021 St.~Petersburg, Russia}
	\affiliation{$^3$Peter Gr\"unberg Institute (PGI) 		
		\& J\"ulich Aachen Research Alliance (JARA-FIT)
		52425 J\"ulich, Germany
	}
\begin{abstract}	
		We report on the observation of a terahertz radiation induced photon drag effect in epitaxially grown \textit{n}- and \textit{p}-type (Bi$_{1-x}$Sb$_{x}$)$_{2}$Te$_{3}$ three dimensional topological insulators with different antimony concentrations $x$ varying from 0 to 1. We demonstrate that the excitation with polarized terahertz radiation results in a \textit{dc} electric photocurrent.  While at normal incidence a current arises due to the photogalvanic effect in the surface states, at oblique incidence it is outweighed by the trigonal photon drag effect. The developed microscopic model and theory show that the photon drag photocurrent is due to the dynamical momentum alignment by time and space dependent radiation electric field and implies the radiation induced asymmetric scattering in the electron momentum space.
\end{abstract}	
	\maketitle{} 
	
\section{Introduction}

Much attention in condensed-matter physics is currently directed towards understanding electronic properties of Dirac fermions in three dimensional (3D) topological insulators (TIs), which challenge fundamental concepts and hold a great potential for electronic, optic and optoelectronic applications, see  e.g. Ref.~\cite{HasanKane2010,Moore2010,QiZhang2011,newreview,book2013,book2013b,book2013c,book2015,book2015b}.  

Recently nonlinear high frequency electron transport phenomena~\cite{Ganichev2003,Ivchenko2008,book_Ivchenko,GlazovGanichev_review} in TI systems have attracted growing interest.
There have been many theoretical and experimental works in the past few years on the helicity controlled photocurrents~\cite{Hosur2011,McIver2012_2,Junck2013,Duan2014,Shikin2015}, linear photogalvanic effect~\cite{Olbrich2014,Braun2015,Zhu2015}, local photocurrents~\cite{Kastl2012,Kastl2015_1}, edge photocurrents in 2D TI~\cite{Kvon2014,Kaladzhyan2015}, coherent control of injection currents~\cite{Bas2012,Muniz2014}, photon drag currents~\cite{Olbrich2014,Tanaka2014}, second harmonic generation~\cite{McIver2012}, photo-induced quantum hall insulators~\cite{Kitagawa2011,Dora2012}, cyclotron resonance assisted photocurrents \cite{Olbrich2013,Dantscher2015}, quantum oscillations of photocurrents\cite{Zoth2014},  photogalvanic currents via proximity interactions with magnetic materials~\cite{Yao2012,Semenov2012,Li2014} and  photoelectromagnetic effect~\cite{Egorova2015}. These phenomena, scaling in the second or third order of the radiation electric fields, open up new opportunities to the study of Dirac fermions, as it has been already demonstrated for graphene, for review see Ref.~\onlinecite{GlazovGanichev_review}, and several TI materials, see e.g. Ref.~\onlinecite{McIver2012_2,Olbrich2014,Kastl2015_1,Dantscher2015}. An important advantage of the nonlinear high frequency transport effects is that some of them, being forbidden by symmetry in the bulk of most 3D TI, can be applied to selectively probe the surface states even in TI materials with a finite bulk conductivity. Utilizing photocurrents this advantage has been used to study Sb$_2$Te$_3$ and Bi$_2$Te$_3$ 3D TIs~\cite{Olbrich2014}, in which conventional \textit{dc} transport experiments, particularly at room temperature, are handicapped by a large residual bulk charge carrier density~\cite{Checkelsky2009,TaskinAndo2009,Analytis2010_PRB, Ren2010, Qu2010, Barreto2014}. 

It has been shown in Ref.~\onlinecite{Olbrich2014} that the photocurrent excited by normal incident terahertz (THz) radiation is generated due to the photogalvanic effect. The latter originated from the 
asymmetric scattering of Dirac fermions driven back and forth by the \textit{ac} electric field and is allowed in the non-centrosymmetric surface states only.  The experiments further hinted at a possible contribution of the photon drag effect - a competing photocurrent resulting from the light momentum transfer to charged carriers. However, no experiments which provide clear evidence of the photon drag effect in TI materials have been reported so far. 
	
Here we report on the observation of the photon drag effect in (Bi$_{1-x}$Sb$_{x}$)$_{2}$Te$_{3}$  3D TIs excited by THz radiation. We demonstrate that while at normal incidence the photocurrent is dominated by the photogalvanic effect, at oblique incidence it is outweighed by the photon drag effect. The latter  is shown to be caused by the \textit{in-plane} component of the photon wavevector ${\bm q}_\parallel$. Strikingly,  the observed photon drag current does not change its sign upon inverting ${\bm q}_\parallel$.  This, seemingly surprising result, is caused by the fact that in materials with trigonal symmetry the photon drag current is proportional not only to the photon wavevector but also to the product of \textit{in-} and \textit{out-of-plane} components of the radiation electric field. Since both, ${\bm q}_\parallel$ and the product of the electric fields, change their sign, the total sign remains unchanged. Our experimental findings are well described by the developed theory and microscopic model, based on the Boltzmann kinetic equation for the carrier distribution function. Both photon drag and photogalvanic effects are investigated in epitaxially grown (Bi$_{1-x}$Sb$_{x}$)$_{2}$Te$_{3}$ bulk materials of various composition determined by the antimony content
$x$. The variation of $x$ enabled us to study photocurrents in different systems including binary and ternary TIs
with smooth changes from $n$- to $p$-type bulk conductivity, see Ref.~\onlinecite{ref-ternaries-Zhang,ref-ternaries} as well as in heterostructure samples, consisting of a $n$-type Bi$_2$Te$_3$ and a $p$-type Sb$_2$Te$_3$ layer, see Ref.~\onlinecite{ref-pn}. In the latter the chemical potential can be tuned by varying the thickness of the upper Sb$_2$Te$_3$ layer.

\section{Sample Description}
\label{section_samples}

\begin{figure*}
	\includegraphics[width=0.9\linewidth]{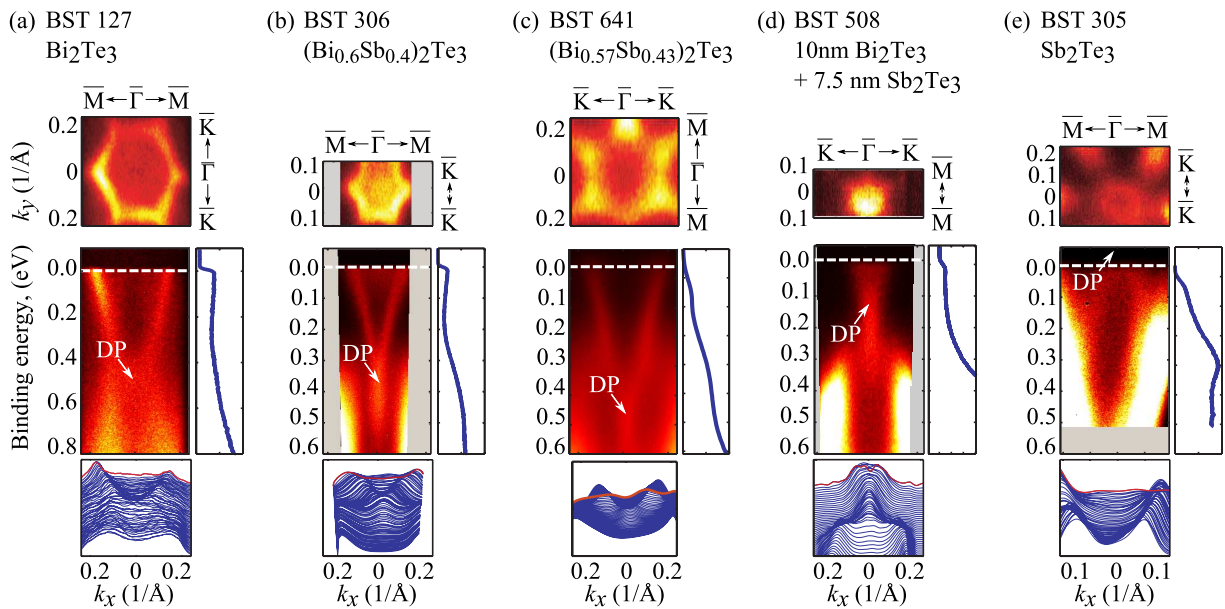}
	\caption{ 
		ARPES investigation of the surface electronic structure of different TI samples, measured at low temperature ($T\approx 25$~K) using photon energy $8.4$~eV. The spectra unambiguously proof the existence of topological surface states in each material. Panel (a) shows results for pure, $n$-type Bi$_2$Te$_3$. Panel (b) and (c) show the corresponding results for samples BST306 and BST641, i.e. the $n$-type (Bi$_{0.6}$Sb$_{0.4}$)$_2$Te$_3$ and (Bi$_{0.57}$Sb$_{0.43}$)$_2$Te$_3$ ternary TIs, (d) for sample BST508, i.e. the $n$-type 10~nm Bi$_2$Te$_3$/ 7.5~nm Sb$_2$Te$_3$ TI heterostructure, as well as (e) for sample BST305, i.e. the $p$-type Sb$_2$Te$_3$. 
		The top panels depict the constant energy contour at $\varepsilon_B = 0$ with indicated crystallographic directions showing the hexagonal warping of energy spectrum of the topological surface states. The middle panel illustrates the binding energy dispersion spectra $\varepsilon(k)$ map at $k_{y} = 0$ along $\bar{\Gamma}M$ direction where the upper part of the topological surface states is revealed while the Dirac point is buried in the bulk valence band maximum. Additionally, the energy distribution curves, integrated over the entire image are shown aside the middle panels. The bottom panels depict the respective momentum distribution curves. The ones for $\varepsilon_B = 0$ are highlighted red color.
	}
	\label{arpes}
\end{figure*}	

\begin{table*}
	\begin{tabular}{|l|l|l|l|l|l|l|}	
		\hline		
		Sample ID	& Type 				& Structure 								& Sb content	& Type~Bulk		&$A_x$~(nA~cm$^2$/W) 	&$A_x$~(nA~cm$^2$/W)	\\  	
		&					&											& $x$			& carriers		&$\theta=0$				&$\theta=180^{\circ}$ 	\\ 		
		\hline 
		\hline	
		BST~127		& binary			& 20 ~nm~ Bi$_{2}$Te$_{3}$					& 0 			& $n$			&$0.32$					&$0.8$					\\
		\hline
		BST~307		& binary			& 13 ~nm~ Bi$_{2}$Te$_{3}$					& 0				& $n$			&$3.1$					&$7$					\\		
		\hline
		BST~323		& ternary			& 24 ~nm~ (Bi$_{1-x}$Sb$_{x})_{2}$Te$_{3}$ 	& 25			& $n$			&$1.8$					& -						\\
		\hline
		BST~306		& ternary			& 23 ~nm~ (Bi$_{1-x}$Sb$_{x})_{2}$Te$_{3}$	& 40			& $n$			&$0.06$					&$0.04$					\\
		\hline
		BST~641		& ternary			& 175~nm~ (Bi$_{1-x}$Sb$_{x})_{2}$Te$_{3}$	& 43 			& $n$			&$0.4$					&$0.6$					\\ 
		\hline
		BST~305		& binary			& 27 ~nm~ Sb$_{2}$Te$_{3}$					& 100 			& $p$			&$0.025$				&$0.02$					\\	
		\hline	
		BST~508		& heterostructure	& 10 ~nm~ Bi$_{2}$Te$_{3}$					& pn 			& $n$			&$0.3$					& -			 			\\ 	
		& 					& ~$+$~7.5~nm~Sb$_{2}$Te$_{3}$				&				&				&						&						\\ 
		\hline			
	\end{tabular}
	\caption{Sample parameters and the amplitudes of photocurrents $A_x$ excited by normal incident radiation with {$f = 3.3$~THz}. Angles of incidence $\theta=0$ and 180$^\circ$ correspond to the front and back excitation, respectively.}
	\label{table}
\end{table*}
	
The samples were grown by molecular-beam epitaxy (MBE) on Si(111) substrates in the so-called van der Waals (vdW) growth mode~\cite{vdW}, i.e. there are only weak bonds between substrate and the TI epilayers, so that the large lattice mismatch does not hinder to grow single crystal TI films with a high structural quality~\cite{Borisova2012,Borisova2013,Plucinski2013}. 
Before insertion into the MBE chamber, the Si(111) surface was chemically cleaned to remove the native SiO$_2$ and to passivate the surface with hydrogen. Prior to the TI layer deposition the substrate was heated up to $600{^\circ}$C for 20~min to desorb the hydrogen atoms. The Bi, Sb and Te atoms were deposited on the substrate using effusions cells, working at temperatures of 530${^\circ}$C (Bi), 450${^\circ}$C (Sb) and 380${^\circ}$C (Te), whereas the substrate temperature was 300${^\circ}$C. The Bi$_{2}$Te$_{3}$ (Sb$_{2}$Te$_{3}$) layer were deposited with a slow growth rate of 27~nm/h (9~nm/h). The sample BST641 was grown at temperature $T_{\rm Bi} = 470^\circ$C, $T_{\rm Sb} = 417^\circ$C, $T_{\rm Te} = 330^\circ$C, 
$T_{\rm substrate} = 300^\circ$C, and with a growth rate of 10 nm/h for 1050 minutes, which corresponds to the thickness of 175~nm. The structure composition and thickness of all investigated samples are given in the Table~\ref{table}. To characterize the samples, electrical measurements on Hall~-~bar structures using standard four point probe and lock-in technique were carried out at $T=1.5$~K.
The bulk charge carrier densities have been determined as following: For binary $n$-type Bi$_2$Te$_3$ and $p$-type Sb$_2$Te$_3$ the bulk carrier density is $n\approx p\approx 5 \times 10^{19}$~cm$^{-3}$. For ternary (Bi$_{1-x}$Sb$_{x}$)$_{2}$Te$_{3}$ alloys with $x$ of about 0.4~-~0.5 we obtain $n\approx p\approx 5 \times 10^{18}$~cm$^{-3}$. The sample (Bi$_{0.57}$Sb$_{0.43}$)$_{2}$Te$_{3}$ (BST641), most insulating in the bulk, is $n$-type and has a carrier density of $n=3 \times 10^{17}$~cm$^{-3}$.

The existence of topologically protected surface states has been verified by means of angle\,-\,resolved\,-\,photoemission\,-\,spectroscopy (ARPES)~\cite{ARPES_2,ARPES_3}, see Fig.~\ref{arpes}. Selected samples have been exposed to air and were transferred into a laboratory-based high-resolution ARPES chamber. In order to remove the oxidized layer and surface contaminants, the samples needed to be cleaned by repeated steps of gentle sputtering, using 750~eV Ar ions, and annealing to $250^\circ$~C - $280^\circ$~C. After this cleaning procedure ARPES maps have been obtained at low temperatures ($T\approx25$~K) employing a monochromatized microwave-driven Xe source with a photon energy of 8.4~eV. Topological surface states have been identified in all of the samples, see also Refs.~\onlinecite{ref-ternaries,ref-pn}. Further, the energetic position of the Dirac point $\varepsilon_B(DP)$ was extracted from the ARPES data. For all samples $\varepsilon_B(DP)$ are in the order of hundreds of meV being comparable with values reported earlier for similar materials~\cite{ref-ternaries-Zhang}.

Additionally, X-Ray Diffraction (XRD) measurements were performed in order to verify the single-crystallinity of the thin films, the orientation with respect to the Si(111) substrate and to determine the domain orientation, see Fig.~\ref{xrd}. The XRD data demonstrate the formation of two trigonal domains, being mirror-symmetric to each other and show, however, that the majority of the domains have the same orientation~\cite{Olbrich2014,Kampmeier2015}. The domains can also be seen in the atomic force microscopy images showing trigonal islands (not shown). 
Height profiles prove that the quintuple layers (QL) have steps of about 1\,nm~\cite{Olbrich2014,Borisova2013}. Using the XRD results we prepared squared shaped samples with edges cut along crystallographic axes $x$ and $y$, see Fig.~\ref{xrd}. To enable electrical measurements two pairs of ohmic contacts have been prepared in the middle of the $5\times5$\,mm$^2$ squared sample's edges.	
			
\section{Experimental Technique}
\label{section_technique}

\begin{table}
	\begin{tabular}{|p{1.5cm}|p{0.8cm}|p{0.8cm}|p{0.8cm}|p{0.8cm}|p{0.8cm}|p{0.8cm}|}
		\hline
		$f$ (THz)			& 3.9 	& 3.3 	& 2.0 	& 1.1 	& 0.8  & 0.6	\\
		\hline
		$\lambda$ ($\mu$m) 	& 77 	& 90	& 148	& 280 	& 385  & 496	\\
		\hline
	\end{tabular}
	\caption[Table caption text]{Frequencies and corresponding wavelengths used in the experiments.}
	\label{table2}
\end{table}	

Experiments on photocurrents in (Bi$_{1-x}$Sb$_{x}$)$_2$Te$_3$ 3D TIs were performed applying radiation of a high power pulsed molecular THz laser~\cite{book}. Using NH$_3$, D$_2$O and CH$_3$F as active gases for the optically pumped laser, 40~ns pulses with peak power of $P \approx $ 10~kW were obtained at different frequencies $f$, see Table~\ref{table2} and 
Refs.~\onlinecite{Tunnel1993,DX1995,Tunnelingfrequency1998}. 
The radiation induces indirect (Drude-like) optical transitions, because the photon energies are much smaller than the carrier Fermi energy. The beam had an almost Gaussian form, which was measured by a pyroelectric camera~\cite{Ganichev1999,Ziemann2000}. A typical spot diameter depends on the  radiation frequency and varies between 1 and 3~mm. The electric field amplitude $E_0$ of the incoming radiation was varied from about $1$ to $30$~kV/cm (radiation intensities $I$ from about $1$ to 1000~kW/cm$^2$).
	
The samples were illuminated at normal and oblique incidence. In experiments at normal incidence front and back illumination was used with angle of incidence $\theta=0$ and $180^\circ$, respectively, see Figs.~\ref{PGE} and ~\ref{qz}. In the measurements applying oblique incident radiation the angle $\theta$ was varied between $-35^\circ$ and $35^\circ$, see insets in Figs.~\ref{theta_alpha} and~\ref{theta}. Note, that larger angles of incidence were not used in order to avoid the illumination of contacts and edges. The photocurrents were analysed in two directions, $x$ and $y$, perpendicular to each other and parallel to the sample edges, see inset in Fig.~\ref{PGE}.  Most experiments at oblique incidence were carried out for ($yz$) plane of incidence. In some additional measurements the orientation of the plane of incidence was rotated by the angle $\psi$ in respect to the ($yz$) plane, see inset in Figs.~\ref{psi_alpha} and~\ref{psi}. 
	
The \textit{dc} photocurrent $J$ was excited in the temperature range from $T=4.2$ to 296~K. It was measured as a voltage drop, $U \propto J$, across a 50\,$\Omega$ load resistor and recorded in unbiased samples with a storage oscilloscope. To control the incidence power of the laser the signal was simultaneously measured at a reference THz detector~\cite{Ganichev84p20}. To examine the photocurrent behavior upon the variation of the polarization state half and quarter wavelength plates were employed. The initial laser radiation was linearly polarized along the $y$-axis. By using $\lambda/2$ plates, the azimuth angle $\alpha$ was varied between the linear polarization on the sample and the $y$ axis, see inset and top panel in Fig.~\ref{PGE}. 

By applying $\lambda$/4 plates, we obtained elliptically (and circularly) polarized radiation. In this case, the polarization state is determined by the angle $\varphi$ between the plate optical axis and the incoming laser polarization. 
Here, the electric field vector is lying parallel to the $x$ axis. 
The polarization states for several $\varphi$ are shown on the top panel of Fig.~\ref{phi}. In this geometry, the radiation helicity is varied as $P_{\rm circ} = \sin{2 \varphi}$~\cite{book,BelkovSSTlateral}.
		
\section{Experimental Results}
\label{section_results}

\begin{figure}
	\includegraphics[width=0.9\linewidth]{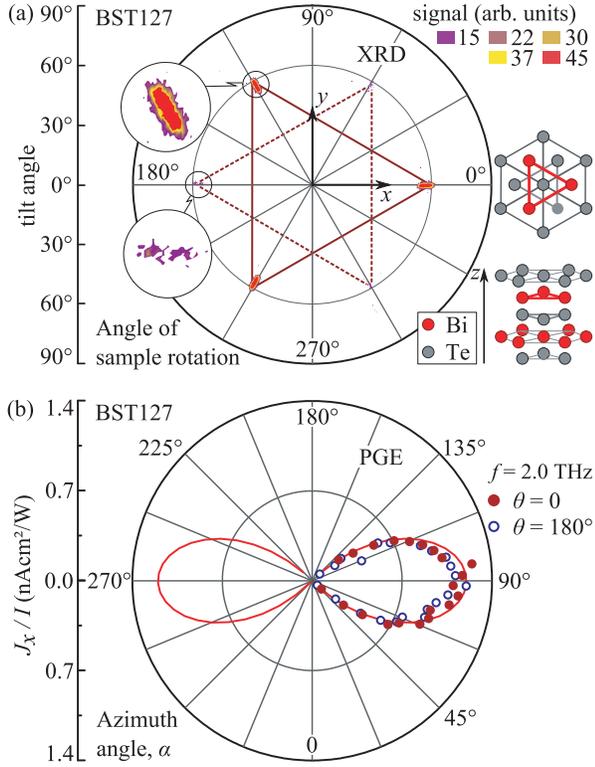}
	\caption{
		(a) $X$-ray diffraction pole figure scan around the (1,0,5) reflection of the Bi$_2$Te$_3$ sample BST127 showing that one domain orientation (highlighted by solid red line) dominates. It also reveals that the crystallographic axes lie parallel to the sample edges. Insets sketches domain orientations illustrated by solid red line connecting top Bi atoms in the upper right panel and side view of one quintuple layer, see right bottom panel. 
		(b) Photocurrent $J_x(\alpha)/I$ in Bi$_2$Te$_3$ sample BST127 measured for front and back illumination at $T=296$\,K. 
		Solid lines show fits after Eq.~\eqref{fit_alpha}. Note that the same dependencies are obtained after phenomenological, see Eq.~\eqref{yz1}, and microscopic theory, see Eq.~\eqref{PGE_final} for the photogalvanic effect.
	}
	\label{xrd}
\end{figure}

\begin{figure}
	\includegraphics[width=0.9\linewidth]{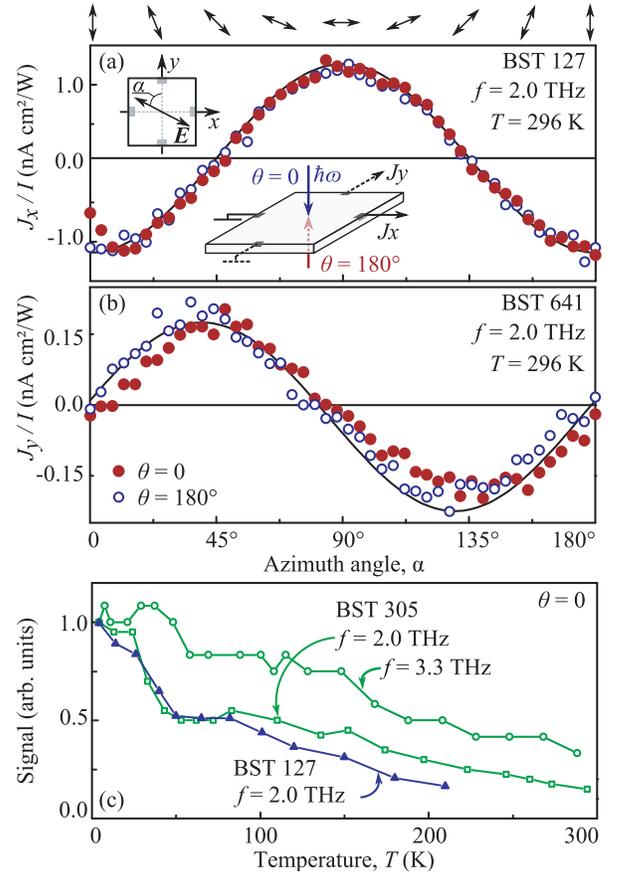}
	\caption{
		(a) Photocurrent $J_x/I$ measured in Bi$_2$Te$_3$ sample BST127.
		(b) Photocurrent $J_y/I$ measured in (Bi$_{0.57}$Sb$_{0.43}$)$_2$Te$_3$ sample BST641. 	
		Plots show the dependence of the photocurrent excited by normal incident radiation with $f = 2.0$~THz on the azimuth angle $\alpha$. Angles of incidence $\theta=0$ and 180$^\circ$ correspond to the front and back excitation, respectively. Solid lines show fits after Eq.~\eqref{fit_alpha}. Note that the same dependencies are obtained after phenomenological, see Eq.~\eqref{yz1}, and microscopic theory, see Eqs.~\eqref{PGE_final} and 26 for the photogalvanic and the photon drag effect, respectively. Insets sketch the setup and the orientation of the electric field. Note, that the photocurrent is probed in the directions coinciding with the principal axes of the trigonal system.  
		(c) Temperature dependence of the photocurrent measured in the Bi$_2$Te$_3$ sample BST127 and Sb$_2$Te$_3$ sample BST305. All data are normalized to the value for $T=4.2$~K.}
	\label{PGE}
\end{figure}

Irradiating the (Bi$_{1-x}$Sb$_{x}$)$_2$Te$_3$ 3D TIs with linearly polarized THz radiation we observed a \textit{dc} current in both $x$- and $y$-~direction. 
The photocurrent was detected in the whole frequency range used from 0.6 up to 3.9~THz. The signal followed the temporal structure of the laser pulse. Its variation upon rotation of the polarization plane is well fitted by
\begin{align}
\label{fit_alpha}
J_x = [- A(f) \cos 2\alpha + C(f)] E_0^2 \, , \\
J_y = [ A(f) \sin 2\alpha + C'(f)] E_0^2 \, ,\nonumber
\end{align}
in which $E_0^2 \propto I$ is the squared radiation electric field and $A$, $C$, and $C'$ are fitting parameters, see Fig.~\ref{PGE} (a) and (b). The above polarization dependencies were observed in all samples and for all frequencies. Cooling the sample from room temperature to 4.2~K did increase the photocurrent amplitude, see Fig.~\ref{PGE} (c), whereas the overall behavior remained unchanged.

\begin{figure}
	\includegraphics[width=0.9\linewidth]{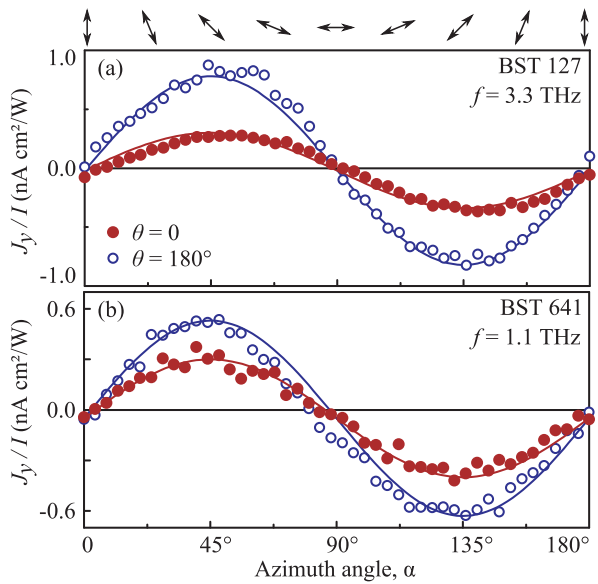}
	\caption{(a) $J_y/I$ measured in Bi$_2$Te$_3$ sample BST127 for $f = 3.3$~THz.   
		(b) $J_y/I$ measured in (Bi$_{0.57}$Sb$_{0.43}$)$_2$Te$_3$ sample BST641 for $f = 1.1$~THz.
		The data show the dependence of the photocurrent  on the azimuth angle $\alpha$ excited by normal incident radiation. Angles of incidence $\theta=0$ and 180$^\circ$ correspond to the front and back excitation, respectively. Solid lines show fits after Eq.~\eqref{fit_alpha}. Note that the same dependencies are obtained after phenomenological, see Eq.~\eqref{yz1}, and microscopic theory, see Eqs.~\eqref{PGE_final} and~\eqref{PDqz_final} for the photogalvanic and the $q_z$~-~photon drag effects, respectively.
	}
	\label{qz}
\end{figure}

As we have shown in Ref.~\onlinecite{Olbrich2014} two phenomena can be the cause of the THz radiation induced photocurrents, described by Eqs.~(\ref{fit_alpha}), namely the photogalvanic and the photon drag effects~\cite{footnote1}. Experiments applying front and back illumination with normally incident radiation allows us to distinguish them from each other. While the photogalvanic current is determined by the \textit{in-plane} orientation of the radiation electric field~\cite{Olbrich2014} and, consequently, remains unchanged for both geometries, the photon drag current is additionally proportional to a component of the photon momentum ${\bm q}$.  Therefore, changing ${\bm q} \rightarrow - {\bm q}$  (front to back illumination) does not affect the photogalvanic but inverts the sign of the factor $A(f)$ for the photon drag effect. Note that for front and back illumination at normal incidence the wavevector $\bm q$ is directed parallel or anti-parallel to the $z$ direction. As an important result we obtained that the sign of the amplitude $A(f)$ remains unchanged (see exemplary Figs.~\ref{PGE} and~\ref{qz} for samples BST127 and BST641 and Table~\ref{table} for all investigated samples at $f = 3.3$~THz). This fact provides a clear evidence that the photocurrent at normal incidence is dominated by the photogalvanic effect in the investigated two dimensional (2D) Dirac fermion systems. In samples BST127 and BST641 excited with 
$f=2.0$~THz, the contribution of the photon drag is 
vanishingly small, see Fig.~\ref{PGE}. 
At other frequencies and samples the photon drag effect
may yield a contribution up to one third compared to that of the photogalvanic effect resulting in larger signals for back illumination than 
front one~\cite{footnote2}, see Fig.~\ref{qz} and Table~\ref{table}.

\begin{figure}
	\includegraphics[width=0.9\linewidth]{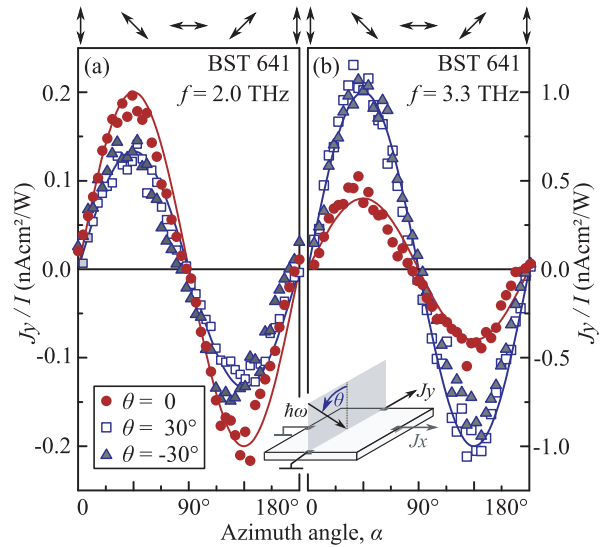}
	\caption{Azimuth angle dependencies of the photocurrent $J_y/I$ excited in (Bi$_{0.57}$Sb$_{0.43}$)$_2$Te$_3$ sample BST641 by normal and oblique incident radiation. The data demonstrate that  the polarization dependence does not change upon variation of the angle of incidence: neither for the case that the current decrease at oblique incidence [panel (a), $f = 2.0$~THz] nor for the case it increases with $|\theta|$ increasing [panel (b), $f = 3.3$~THz]. 
	Solid lines show fits after Eq.~\eqref{fit_alpha}. Note that the same dependencies are obtained after phenomenological, see Eq.~\eqref{yz1}, and microscopic theory, see Eqs.~\eqref{PDqx_final},~\eqref{PGE_final} and~\eqref{PDqz_final} for the photogalvanic and both photon drag effects.
	}
	\label{theta_alpha}
\end{figure}

\begin{figure*}
	\includegraphics[width=0.9\linewidth]{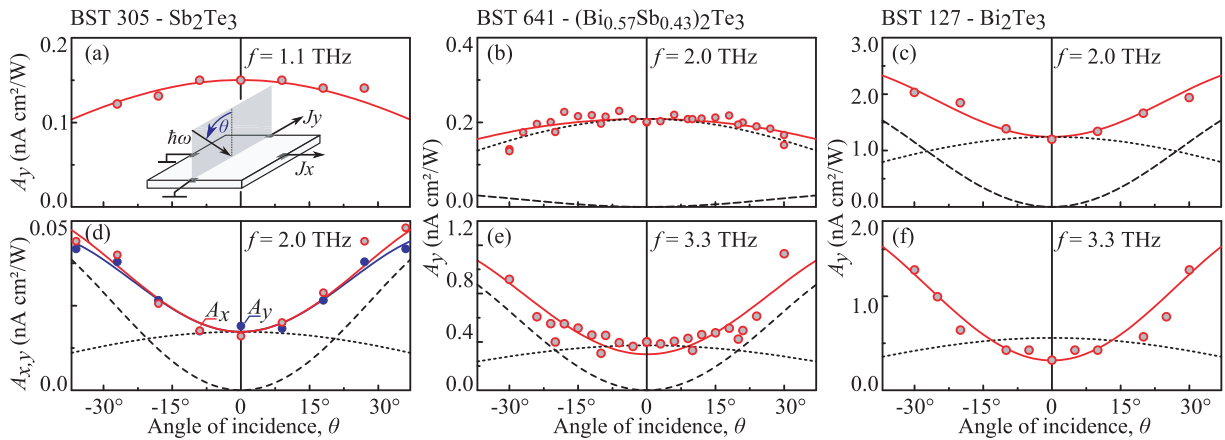}
	\caption{Dependencies of the photocurrent amplitudes $A_{x,y}$ on the angle of incidence $\theta$ obtained for different frequencies and three samples: Sb$_2$Te$_3$ (BST305), (Bi$_{0.57}$Sb$_{0.43}$)$_2$Te$_3$ (BST641) and Bi$_2$Te$_3$ (BST127). Solid lines show fits after see Eq.~\eqref{yz1} and calculations after equations~\eqref{PGE_final},~\eqref{PDqz_final} and~\eqref{PDqx_final}. Dotted  and dashed lines show  contributions of the photogalvanic effect and photon drag effect caused by the $q_x$ component of the photon wavevector. The curves are calculated after Eqs.~\eqref{PGE_final} and after Eqs.~\eqref{PDqx_final}, respectively. Note that solid curves in panel (e) and (f) are obtained also taking into account contribution of the photon drag effects caused by $q_z$ component of the photon wavevector. The latter effect (not shown) mainly contributes to the signal at normal incidence and is the cause for the difference between the value of the calculated total photocurrent (solid line) and the photogalvanic effect contribution (dotted line). It has  a negative sign and decreases with the $\theta$ increasing. The relative contributions of the photogalvanic and photon drag effects are obtained from the measurements applying front and back illuminations, see Figs.~\ref{PGE} and~\ref{qz} as well as Table~\ref{table}.  Inset in panel (a) sketches the setup} 
	\label{theta}
\end{figure*}

So far we presented data obtained for normal incidence. Illuminating the samples at oblique incidence, we found that all characteristic properties of the photocurrent including its polarization behavior remain unchanged, see Fig.~\ref{theta_alpha}. In contrast to the measurements at normal incidence, the magnitude $A_{x,y}(f,\theta)$ depends now additionally on the incident angle $\theta$, as well as on the direction in which the current is examined:  at large $\theta$ 	and for the plane of incidence coinciding, e.g. with ($yz$)-plane the photocurrent magnitudes measured in $x$ and $y$ directions become slightly different. For some excitation frequencies the photocurrent amplitude $A_{x,y}(f,\theta)$ did reduce upon the increase of the angle $\theta$, exemplarily shown for the current measured in $x$ direction in Fig.~\ref{theta_alpha} (a), Fig.~\ref{theta} (a) and (b). Note that the change of $A_{x,y}(f,\theta)$ is even in the angle $\theta$.
	
\begin{figure}
	\includegraphics[width=0.9\linewidth]{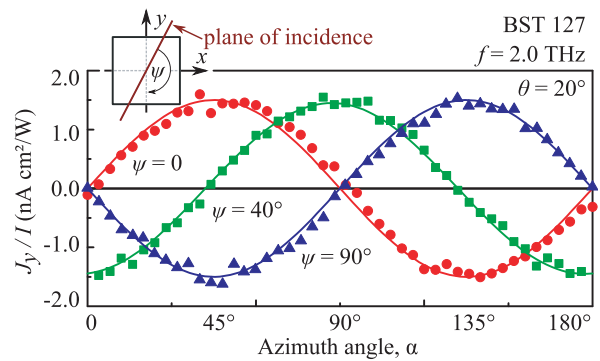}
	\caption{Azimuth angle dependencies of the photocurrent $J_y/I$ excited in Bi$_2$Te$_3$ sample BST127. The data are obtained at oblique incidence radiation ($f = 2.0$~THz) for $\theta = 20^\circ$ and different orientations of the plane of incidence in respect to the $y$-direction given by the angle $\psi$, see inset. The data are shown for the angles $\psi=0$, $40^\circ$ and $90^\circ$. The solid lines are calculated after Eq.~\eqref{yz1o}.
	}
	\label{psi_alpha}
\end{figure}

\begin{figure}
	\includegraphics[width=0.9\linewidth]{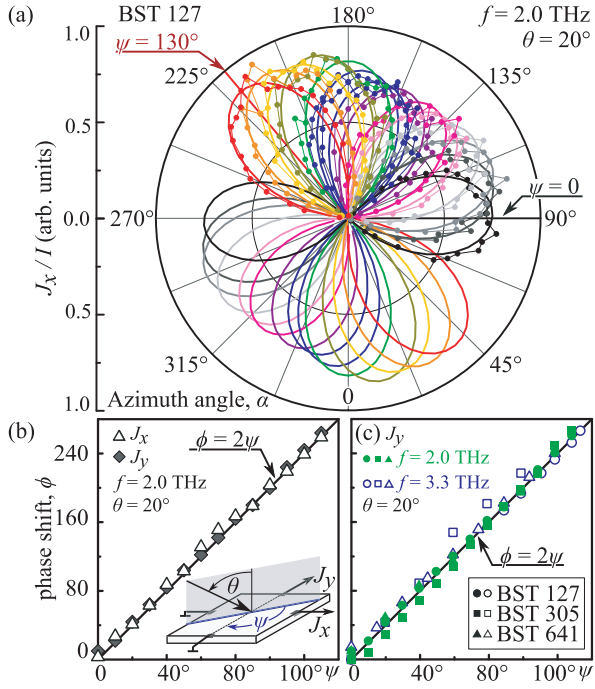}
	\caption{(a) Azimuth angle dependencies of the photocurrent $J_x/I$ excited in Bi$_2$Te$_3$ sample BST127. The data are obtained at oblique incidence radiation ($f = 2.0$~THz) for $\theta = 20^\circ$ and different orientations of the plane of incidence in respect to the $y$-direction given by the angle $\psi$. The data are shown for the angle $\psi$ changing from  $0^\circ$ to $130^\circ$ with steps of $\Delta \psi = 10^\circ$.  Solid lines show fits after Eqs.~\eqref{yz1o}. Bottom panels show the measured phase shift as a function of $\psi$  for (b) the data shown in panel (a) and (c)  three samples BST641, BST127 and BST305 and two radiation frequencies.  Here solid and open symbols correspond to the frequencies $f = 2.0$ and 3.3~THz, respectively. Solid line shows fit after $\phi = 2\psi$.}
	\label{psi}
\end{figure}

Strikingly, at other radiation frequencies we observed that for positive as well as for negative $\theta$ the signal rises with an increase of the angle of incidence, see Fig.~\ref{theta_alpha} (b) and Fig.~\ref{theta} (c)~-~(f). This behavior is observed for photocurrents measured in directions parallel  as well as normal to the plane of incidence, see Fig.~\ref{theta} (d). It is also detected for any orientation of the plane of incidence.  Figure~\ref{psi_alpha} shows the corresponding data for three positions of the incident plane determined by the angle $\psi$. The figure reveals that the signal varies as $J_{y} = A_{y}(f,\psi) \sin (2 \alpha - \phi)E_0^2$ and the most pronounced change in the photocurrents' polarization dependence is the appearance of a $\phi = 2\psi$ phase shift. A detected small variation of $A_y(f,\psi)$ as a function of $\psi$  can not be discussed earnestly, since precise adjustment ensuring that for different $\psi$, technically obtained by rotating the sample, the laser spot remains on the same sample position is hard to realize. To support the conclusion that the rotation of the incident plane results in a $2\psi$ phase shift, we measured photocurrents depending on the linear polarization by changing the angle $\psi$ by steps of~$10^\circ$ in the range from 0 to $130^\circ$. Figure~\ref{psi} (a) demonstrates the dependencies obtained for Bi$_2$Te$_3$ sample BST127. The corresponding dependency of the measured phase shift $\phi$ on the angle $\psi$  is shown in Fig.~\ref{psi} (b) and for other samples and frequencies in ~Fig.~\ref{psi} (c). The figures demonstrate that in all cases $\phi \approx 2\psi$.
	
\begin{figure}
	\includegraphics[width=\linewidth]{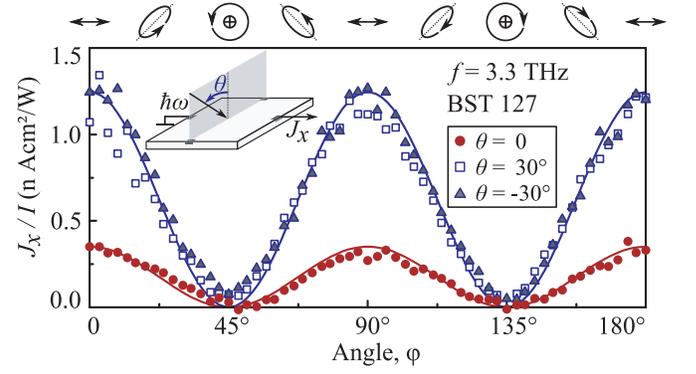}
	\caption{
		Helicity dependence of the photocurrent, $J_x/I$, measured  in Bi$_2$Te$_3$ sample BST127 at normal as well as at oblique incidence in the direction normal to the plane of incidence. The ellipses on top illustrate the polarization states for various angles $\varphi$. Solid lines show fits after Eqs.~\eqref{yz1} where the polarization dependent terms take for this geometry the form $J_x = A_x(f)(\cos 4 \varphi + 1)/2$.}
	\label{phi}
\end{figure}

Finally, we discuss the results obtained applying elliptically (circularly) polarized radiation. These measurements were particularly motivated by the search for the circular photogalvanic~\cite{Resonantinversion2003,SGEopt2003,Hosur2011} and circular photon drag effect~\cite{Diehlcircdrag2007,Karch2010}, i.e. photocurrents changing their direction upon switching of the radiation helicity~\cite{Ganichev2003,Ivchenko2008,GlazovGanichev_review}, recently observed for Bi$_2$Te$_3$ TI excited by near infrared light~\cite{McIver2012}. Applying radiation at oblique incidence and measuring the photocurrent in the direction normal to the plane of incidence ($yz$), i.e. in the geometry for which circular photogalvanic~\cite{McIver2012,3authors,GaN_LPGE} and circular photon drag effects~\cite{GlazovGanichev_review,Diehlcircdrag2007} are expected, we detected a current which can be well fitted by $J_x = A_x(f)(\cos 4 \varphi + 1)/2$, see Fig.~\ref{phi}. The figure clearly shows that for circularly polarized radiation ($\varphi=45^\circ$ and $135^\circ$) the signal vanishes. In fact, the term in brackets describes the degree of linear polarization in the lambda-quarter plate geometry. Therefore, the observed current is identical to the one excited by linearly polarized radiation and which was already discussed above. Examining different samples in the whole investigated THz frequency range, we observed the same result: no trace of a helicity dependent photocurrent has been detected.
	
To summarize, experiments on different types of TI  samples
provide a self-consistent picture demonstrating that the photocurrents 
	(i) are caused by effects proportional to the second power of the radiation electric field and can be excited by both normal and oblique incident radiation, 
	(ii) are excited by linearly polarized radiation, 
	(iii) vary with the azimuth angle $\alpha$ as $J \propto A(f,\theta, \psi) \sin 2 \alpha$ with a possible phase shift depending on the experimental geometry and cristallographic direction in which the photocurrent is measured, see Figs.~\ref{psi_alpha} and~\ref{psi},
	(iv) have the same sign but may have distinct magnitudes of the factor $A(f)$ for front and back normal incident illumination, see~Fig.~\ref{PGE} and~\ref{qz}, and 
	(v) are described by an even function of the angle of incidence $\theta$ with the magnitude $A(f,\theta)$ for different radiation frequencies, raising or decreasing with increase of $\theta$, see Figs.~\ref{theta} and~\ref{theta_alpha}.

\section{Discussion}
\label{section_discussion}
	
Now we discuss the origin of the observed photocurrents, which are induced by spatially homogeneous terahertz radiation and scale with the second power of the radiation electric field.  We begin with the standard way to treat second order effects without going into microscopic details, which makes use of the symmetry arguments. This approach allows us to explore what kind of photocurrents are allowed in the considered system and  to describe their variation  upon change of macroscopic parameters, such as radiation intensity, polarization and incident angle. With this, the response of the charge carriers ensemble to the external field  can be characterized conveniently by the coordinate- and time-dependent electric current density $\bm j(\bm r,t)$.  It is expanded in a power series in the external alternating electric field $\bm E(\bm r,t)$ in the form of a plane wave 
	\begin{equation} 
	\label{plane}
	{\bm E}({\bm r}, t)= {\bm E}(\omega, \bm q) {\rm e}^{- {\rm i} \omega t + {\rm i} {\bm q} {\bm r} } + {\bm E}^*(\omega, \bm q) {\rm e}^{{\rm i} \omega t - {\rm i} {\bm q} {\bm r}}\:,
	\end{equation}
where $\omega = 2\pi f$ is the angular frequency and $\bm q$ is its wavevector. Limiting the consideration to the second order effects  we  obtain the photocurrent density $j \propto J$ in the form~\cite{book_Ivchenko,book}
	\begin{equation}
	\label{Ch7currentgeneral11} 
	j_{\lambda} =\sum_{\mu, \nu} \chi_{\lambda \mu \nu} E_{\mu} E^*_{\nu} + \sum_{\delta, \mu, \nu} T_{\lambda \delta \mu \nu} q_{\delta} E_{\mu} E^*_{\nu} + c.c. \:,
	\end{equation}
where the expansion coefficients $\chi_{\lambda \mu \nu}$ and $T_{\lambda\mu \nu \delta}$ are third and fourth rank tensors, respectively, and $E^*_{\nu} = E^*_{\nu}(\omega ) = E_{\nu}(-\omega)$ is the complex conjugate of $E_{\nu}$. The first term on the right-hand side of Eq.~\eqref{Ch7currentgeneral11} represents photogalvanic effects whereas the second term describes the photon drag effect containing additionally the wavevector of the electromagnetic field.
	
Equation~\eqref{Ch7currentgeneral11} can be simplified considering the point group $C_{3v}$, which describes the symmetry of the surface states in (Bi$_{1-x}$Sb$_{x}$)$_{2}$Te$_{3}$.  To be specific we first obtain the photocurrents excited in the crystallographic directions $x$ and $y$ with the radiation plane of incidence ($yz$). Taking into account the fact that we detected only photocurrents excited by linearly polarized radiation being even in the angle of incidence $\theta$ we can omit all contributions of photocurrents  sensitive to the radiation helicity and those giving the response odd in the angle $\theta$. Under these conditions we derive for $j$
	\begin{align}
	\label{yz1}
	j_x &=  (\chi + T_z q_z) (E_x^2-E_y^2) -  T_\parallel q_y E_y E_z \\
	&= j^{\rm off} -\cos{2\alpha} \, E_0^2\nonumber \\ 
	&\times {1\over 2} \left[ (\chi -T_z q \cos{\theta}) (t_s^2+t_p^2\cos^2{\theta})
	+ 
	T_\parallel q t_p^2\sin^2{\theta} \cos{\theta}\right] , \nonumber\\
	j_y &=  - 2 (\chi + T_z q_z) E_x E_y -  T_\parallel q_yE_xE_z \nonumber \\
	&= \sin{2\alpha} \, t_st_p\left[ (\chi-T_z q \cos{\theta})\cos{\theta} 
	+ T_\parallel q \sin^2{\theta}/2 \right] E_0^2. \nonumber
	\end{align}
Here $t_p$ and $t_s$ are the Fresnel transmission coefficients for $s$- and $p$- polarized light and $j^{\rm off}$ is the polarization independent offset being equal to zero for $\theta = 0$, see Appendix~1. The constants $\chi$, $T_\parallel$ and $T_z$ are coefficients describing the photogalvanic, the photon drag effect at oblique incidence, and the photon drag effect at normal incidence, respectively.
	
At normal incidence we obtain the photogalvanic effect and the photon drag effect, caused by $z$-component of the photon wavevector. They are given, respectively, by the terms proportional to factors $\chi$ and $T_z$. Both effects have the same polarization dependence and vary with the azimuth angle $\alpha$ according to $j_x \propto (E_x^2-E_y^2) \propto \cos{2\alpha}$ and $j_y\propto 2E_x E_y \propto \sin 2\alpha$, being in agreement with the results of experiments, see Figs.~\ref{PGE},~\ref{qz} and Eq.~\eqref{fit_alpha}. For $\chi \geq T q_z$ reversing the radiation propagation direction ($\theta = 0 \Rightarrow \theta = 180^\circ$) does not affect the polarization dependence. It may, however, change the signal magnitude 
because $\chi+T_zq_z$ is different for negative and positive $q_z$ corresponding to front and back illumination.
Exactly this behavior has been observed in experiments showing that at some frequencies the photon drag due to $z$-component of the wavevector yields a minor contribution to the total photocurrent, see values of $J_x$ at $\theta = 0$ and $180^\circ$ in Table~\ref{table} and Fig.~\ref{qz}.  At oblique incidence $E_x$ and $q_z$ components are reduced and the magnitude of both photocurrents  diminish equally for positive and negative angle $\theta$, see Fig.~\ref{theta_alpha} (a) and doted lines in Fig.~\ref{theta}. 

According to Eqs.~\eqref{yz1} at oblique incidence the photon drag effect, coupled to the \textit{in-plane} wavevector $q_y$, can also contribute to the total photocurrent.  Its dependence on the azimuth angle $\alpha$ formally coincides with that of the contributions discussed so far, $j_x \propto E_y E_z \propto \cos 2\alpha$ and $j_y \propto E_xE_z \propto \sin 2\alpha$. However, it obviously vanishes at normal incidence ($q_y = 0$) and, in contrast to photogalvanic effect, increases with the rising  angle of incidence, see dashed lines in Fig.~\ref{theta_alpha}. Moreover, the sign of the products $(E_yE_zq_y)$ and $(E_xE_zq_y)$ is staying the same for positive and negative $\theta$. A dominating contribution of this effect has been observed at large angles of incidence for all samples and for almost all frequencies.  The most clear evidence for this conclusion is supported by measurements shown in Fig.~\ref{theta_alpha} (b) and~\ref{theta} (c)-(f) demonstrate that the photocurrent rises with the angle $\theta$ increase. 
	
Rotation of the incident plane by the angle $\psi$ changes the relative orientation of the electric field and crystallographic axes modifying Eqs.~\eqref{yz1}. Taking into account that in all experiments described above ${T_\parallel q_y \gg (\chi + T_zq_z)}$ and considering small angles $\theta$, being relevant to the experimental data of Figs.~\ref{theta_alpha} and~\ref{theta}, we obtain 
%
	\begin{align}
	\label{yz1o}
	j_x \approx  -\cos{(2\alpha - 2\psi)} &(\chi-T_z q + T_\parallel q \theta^2/2) t^2 E_0^2,\\
	j_y \approx  \sin{(2\alpha - 2\psi)} &(\chi-T_z q + T_\parallel q \theta^2/2) t^2 E_0^2. \nonumber \\ 
	\nonumber
	\end{align}
Here $t=t_p=t_s$ is the amplitude of the transmission coefficient for small $\theta$ and the offset current is omitted. Equations for arbitrary angle of incidence are given in the Appendix~1. Equations ~\eqref{yz1o} show that the rotation of the incident plane mainly results in a $2\psi$ phase shift for both photocurrents $j_x$ and $j_y$. This phase shift has been observed for all samples and frequencies, see Figs.~\ref{psi_alpha} and~\ref{psi}.

\section{Microscopic models}

In general, second order high frequency effects are caused by the redistribution of charge carriers in the momentum space induced by the illumination of the sample with radiation. The resulting nonequilibrium distribution can contain components which are oscillating in time and space, as well as steady-state and spatially homogeneous ones.  Hence, the irradiation may cause both \textit{ac} and \textit{dc} flows in a media. Their magnitudes are nonlinear functions of the field amplitude and their components are sensitive to the radiation polarization. In the following section we present models visualizing the physics of nonlinear responses. For simplicity we assume positively charged carriers, i.e., holes for which the directions of the carrier flow and the corresponding electric current coincide.

\subsection{Trigonal photogalvanic effect}
\label{section_modelpge}

The model and the microscopic theory of the photogalvanic effect have been discussed in detailed in Ref.~\onlinecite{Olbrich2014}, demonstrating that the photocurrent stems from the asymmetric scattering of free carriers excited by irradiation with a \textit{ac} electric field. As we show below the asymmetric scattering is also responsible for the observed photon drag effects. Therefore, to introduce the concepts essential for the formation of the latter effects and to provide a complete picture of the photocurrent formation in TI, we will briefly address the model of the photogalvanic effect. 

	\begin{figure}
		\includegraphics[width=0.9\linewidth]{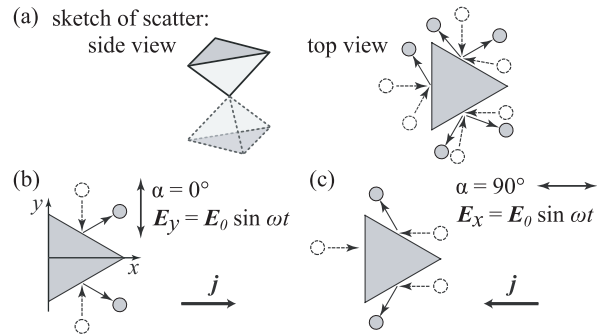}
		\caption{Model of the photogalvanic effect, excited in surface states of (Bi$_{1-x}$Sb$_{x}$)$_2$Te$_3$ due to asymmetry of elastic scattering of holes by wedges.}
		\label{modelPGE}
	\end{figure}	

The current generation process is illustrated in the right panel of Fig.~\ref{modelPGE}.  As addressed above, the symmetry of the surface states in (Bi$_{1-x}$Sb$_{x}$)$_{2}$Te$_{3}$ 3D TIs is C$_{3v}$. This point group implies that the anisotropy of carrier elastic scattering is the same as at scattering by a double triangular pyramid, whose side and top view are sketched in the left panel Fig.~\ref{modelPGE} (a). Note that for C$_{3v}$ symmetry scattering by top and bottom pyramids has  different  probabilities. In the framework of the photogalvanic effect caused by the \textit{in-plane} motion of free carriers the scatterers can be considered as randomly distributed but identically oriented wedges lying in the QL-plane. The preferential orientation of wedges is supported by the $X$-rays data shown above, see Fig.~\ref{xrd} (a) and Ref.~\onlinecite{Olbrich2014,NatPhys09}.  In the absence of radiation, the flows of anisotropically scattered holes, see right panel Fig.~\ref{modelPGE} (a), exactly compensate each other.  Application of linearly polarized THz radiation results in an \textit{alignment} of carrier momenta: the total flow of holes driven back and forth by \textit{ac} electric field $\bm{E}(t)$ increases. The corresponding \textit{stationary} correction to the hole distribution function scales as a square of the $ac$ electric field magnitude~\cite{alignment}. The \textit{stationary alignment} of carrier momenta itself does not lead to a \textit{dc} electric current but, due to asymmetric scattering by wedges, the excess of the flux of carriers moving along the field violates the balance of flows~\cite{Belinicher-Strurman-UFN,GaN_LPGE}, and the linear photogalvanic current is generated~\cite{alignment2}.  The direction of the induced current depends on the relative orientation of the \textit{ac} electric field and wedges: e.g., a field parallel to the wedges base (${\bm E}\parallel y$), see Fig.~\ref{modelPGE}~(b), yields the current flow in $x$-direction while rotation of the electric field by 90$^\circ$ reverses the current direction, see Fig.~\ref{modelPGE}~(c). The polarization dependence of the photogalvanic current in $x$- and $y$- directions is described by the terms with $\chi$ in Eqs.~\eqref{yz1o} and by Eq.~\eqref{PGE_final}. Note that the coefficient $\chi$ has opposite signs for holes and electrons.

\subsection{Trigonal photon drag effects caused by \textit{in-plane} component of the photon wavevector}	
\label{section_modeldrag}

\begin{figure}
		\includegraphics[width=0.9\linewidth]{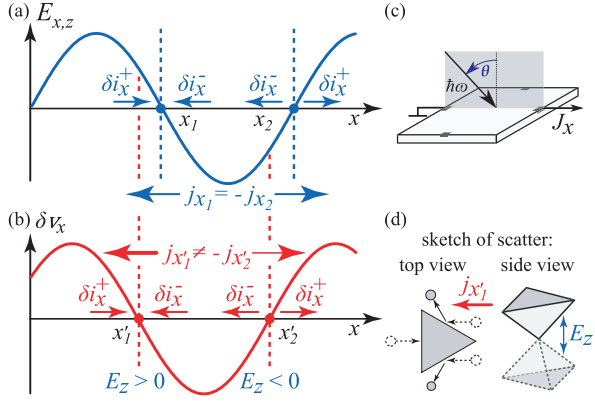}
		\caption{Model of the trigonal photon drag effect caused by the \textit{in-plane} wavevector $q_x$.  To be specific we discuss the hole gas in the surface states excited  
		by oblique radiation with the incidence plane ($xz$), see panel (c). Triangle in (d), left picture, shows top view of the considered scattering potential. Side view of the scatters is sketched in the right picture. }
		\label{modelPD}
	\end{figure}

The trigonal photon drag effect caused by the \textit{in-plane} component of the photon wavevector results in a $dc$ current increasing with the angle of incidence increase.  It is described by an \textit{even} function of the angle $\theta$.  Similar to the photogalvanic effect the photon drag current formation involves asymmetric scattering of free carriers.
The trigonal photon drag effect results from a \textit{dynamical} alignment of carrier momenta. It is generated due to the \textit{in-plane} profile
of the radiation electric field and implies the difference in the scattering probabilities for different half periods of the electromagnetic wave. The process of the current generation is illustrated in Fig.~\ref{modelPD}. Like in the model for photogalvanic effect we consider the scatterers as randomly distributed but identically oriented pyramids in the QL-plane, see Fig.~\ref{modelPD}~(d). 
In the absence of radiation, the flows of the thermalized charge carriers which are anisotropically  scattered by pyramids exactly compensate each other. Optical excitation disturbs the balance due to the action of high frequency electric field $\bm E$ on charged carriers, we assume holes.  The discussed trigonal photon drag current is caused by the dynamic variation of the electric field $\bm E$ in the direction of the radiation propagation, see Fig.~\ref{modelPD}~(a).  The strength of the corresponding force acting on holes is given by $ {|e| \bm E_\parallel \mathrm e^{\mathrm i \bm q_\parallel  \bm r - \mathrm i \omega t}  \approx} \:\, |e| \mathrm i (\bm q_\parallel   \bm r) \bm E_\parallel\mathrm e^{-\mathrm i \omega t}$, where $e$ is the elementary 
charge~\cite{GlazovGanichev_review,perelpinskii73}. 
The force is coordinate-dependent and causes the hole acceleration to be directed parallel to the $x$-direction for $E_x > 0$ (anti-parallel for $E_x < 0$) and, consequently, increase of the hole flow by $\delta i^+_x$ ($\delta i^-_x$), see horizontal arrows in Fig.~\ref{modelPD} (a). As a result of this \textit{dynamical} momentum alignment the balance of the hole flows scattered by pyramids in the vicinity of the electric field $E_x$ nodes becomes \textit{locally} violated, see Fig.~\ref{modelPD} (a) and (d).  The asymmetric scattering may cause equal in magnitude but oppositely directed  local electric currents $j_{x_1}$ and  $j_{x_2}$ in the vicinity of $x_1$ and $x_2$; whereas the total electric current remains zero.  However, steady-state $dc$ electric current indeed emerges if one additionally takes into account the $z$-component of the radiation field and the retardation between the electric field and the instant velocity of charge carrier. The photocurrent reaches its maximum at $\omega \tau$ about unity. The effect of the retardation, well known in the Drude-Lorentz theory of high frequency conductivity, causes a phase shift  between the electric field $E_x$ and the instant change of the hole velocity $\delta v_x$ given by  $\arctan(\omega\tau_{\rm tr})$. 
Consequently, the nodes of the charge carriers velocity $\delta v_x$ are shifted  in respect to that of electric field $E_x$, these nodes are indicated in Fig.~\ref{modelPD} (b) as $x'_1$ and $x'_2$. The carriers in the vicinity of the $x'_1$ and $x'_2$-positions are subjected to the electric field $E_z(x'_1)$ and $E_z(x'_2)$ which have the opposite signs. The $E_z(x'_1)$  [$E_z(x'_2)$] field  is pushing the  carriers to the basis [top] of the pyramids which increase [decrease] the scattering probability  at $x'_1$ [$x'_2$] vicinity. Consequently, it changes the magnitudes of the local currents $j_{x_1'}$ and  $j_{x_2'}$ caused by the  asymmetric scattering. The variation of the scattering probability upon the action of the \textit{out-of-plane} electric field $\delta W_{\bm p' \bm p}(E_z)$ is described by Eq.~\eqref{dW} in Sec.~\ref{sec:micro}. As a result the oppositely directed local  currents 
do not compensate each other anymore and a $dc$ electric current  being proportional to the product $q_x E_x E_z$ emerges. Changing the angle of incidence form $\theta$ to $-\theta$ reverses the sign of both $q_x$ and the product $E_xE_z$ so that  the  direction of the \textit{dc} current remains unchanged. We emphasize that such a contribution to the photon drag effect is specific for trigonal systems and it is absent in, e.g., hexagonal systems like graphene~\cite{GlazovGanichev_review,Karch2010}.

\section{Microscopic Theory}
\label{sec:micro}

Now we turn to the microscopic theory of the photon drag effect. In the classical regime achievable in  our experiments, which is characterized by $\hbar \omega \ll \varepsilon_{\rm F}$, the photocurrents can be well described by means of Boltzmann's kinetic equation for the coordinate dependent carrier distribution function $f_{\bm p}(\bm r)$
\begin{align}
\label{Boltzmann}
	\left({\partial \over \partial t} + e{\bm E}({\bm r}, t) \cdot {\partial \over \partial \bm p} + {\bm v}_{\bm p} \cdot {\partial \over \partial \bm r}\right) f_{\bm p}(\bm r) \\
	= 
	\sum_{\bm p'} [W_{\bm p \bm p'}f_{\bm p'}(\bm r) - W_{\bm p' \bm p} f_{\bm p}(\bm r)]. \nonumber
\end{align}
where $e<0$ for holes and $e<0$ for electrons,  ${\bm v}_{\bm p}=v_0 \bm p/p$ is the velocity of surface charge carriers with a  momentum $\bm p$, $v_0$ is the Dirac fermion velocity, and $W_{\bm p' \bm p}$ is a probability for a charge carrier to have the momenta $\bm p$ and $\bm p'$ before and after scattering, respectively. Lack of inversion center for the surface charge carriers makes their elastic scattering asymmetric, so that ${W_{\bm p \bm p'} \neq W_{-\bm p, -\bm p'} }$~\cite{Belinicher-Strurman-UFN,GaN_LPGE}, and results in a \textit{dc} electric current.  Note that this asymmetry takes place even for isotropic scatterers like impurities or phonons. The photocurrent can be calculated as follows~\cite{book_Ivchenko}
\begin{equation}
\label{j}
	\bm j  = e\sum_{\bm p}  {\bm v}_{\bm p} \delta f_{\bm p} \, ,
\end{equation}
where $\delta f_{\bm p}$ is the correction to the distribution function being quadratic in the radiation electric field amplitude and linear in the photon momentum.

To calculate the asymmetric part of the scattering probability, which is responsible for the photocurrent formation, we take into account warping of the energy spectrum. Without warping the energy dispersion of the surface states is described  by the Hamiltonian~\cite{HasanKane2010}
\begin{equation}
	H_0 = v_0(\sigma_x p_y - \sigma_y p_x) \, ,
	\label{hamiltonian}
\end{equation}
where $\sigma_{x,y}$ are Pauli matrices. The Hamiltonian~\eqref{hamiltonian} yields the linear energy dispersion $\varepsilon_{e,h}=\pm v_0 p$ and corresponding wavefunctions ${\Psi_{e,h}^{(0)} =[1, \mp{\rm i} {\rm exp}{({\rm i}\varphi_{\bm p})}]/\sqrt{2} }$ for electrons and holes, where $\varphi_{\bm p}$ is the angle between carrier momentum $\bm p$ and the $x$ axis. The warping of the energy spectrum reflects the trigonal symmetry of the surface and is described by an additional to Eq.~(\ref{hamiltonian}) small term   given by~\cite{hex_warping_H} 
\begin{equation}
	H_w = \lambda^w \sigma_z p^3 \sin{3\varphi_{\bm p}} \, ,
	\label{warping}
\end{equation}
where $\lambda^w$  is a warping constant. This is the perturbation which leads to the hexagonal warping of the energy surfaces~\cite{lambda2} clearly detected by ARPES, see top panels in Fig.~\ref{arpes}.

The perturbation caused by the terahertz radiation electric field $E_z$ changes the surface charge carrier wavefunctions due to admixture of bulk  states from various bands. The corresponding Hamiltonian is linear in the coordinate $z$
\begin{equation}
H_{em}= -e z E_z .
\end{equation}
Taking into account both perturbations $H_w$ and $H_{em}$ in the first order, we obtain the corrected electron wavefunction:
\begin{equation}
	\Psi_{e}=  \Psi_{e}^{(0)}   + {\lambda^w p^2 \sin{3\varphi_{\bm p}} \over 2v_0}\Psi_{h}^{(0)}  +e E_z  \sum_{n} {z_{ns} \over \varepsilon_n} \Psi_n, \nonumber
\end{equation}
where the index $s$ labels the bulk orbitals from which the surface states are formed, and $n$ enumerates other energy bands of the bulk crystal.
Here we  assume that all bulk bands lie far enough away from the Dirac point so the energies ${|\varepsilon_n| \gg \varepsilon_{\rm F} = v_0p_{\rm F} }$, where $p_{\rm F} $ is the Fermi momentum.

Calculating the matrix elements of scattering by a static potential we obtain from the Fermi golden rule the scattering probability in the form 
$W_{\bm p' \bm p} = W_{\bm p' \bm p}^{(0)} + \delta W_{\bm p' \bm p}$. 
The field-independent part is given by the usual expression taking into account the absence of back scattering for Dirac fermions~\cite{backscattering}.
$$W_{\bm p' \bm p}^{(0)} = {\pi \over \hbar} \left<|V(\bm p' - \bm p)|^2 \right>(1+\cos{\theta_{\bm p' \bm p}})\delta(v_0p-v_0p'),$$ 
where $V(\bm p)$ is the Fourier image of the scattering potential, $\theta_{\bm p' \bm p}=\varphi_{\bm p'}-\varphi_{\bm p}$ is the scattering angle, and the angular brackets mean averaging over  positions of scatterers~\cite{lambda2}. The linear in $E_z$ correction is given by
\begin{align}
\label{dW}
\delta W_{\bm p' \bm p} = & {2\pi \over \hbar }\delta(v_0p-v_0p') 
\sum\limits_{n} \left< {\rm Im} \left(V_{sn} {z_{ns} V_{ss}^*} \right) \right>  /\varepsilon_n
\\
& \times e E_z \sin{\theta_{\bm p' \bm p}}
{\lambda p^2 \over v_0}	(\sin{3\varphi_{\bm p'}} +\sin{3\varphi_{\bm p}}).
\nonumber
\end{align}

Here $V_{ss}$ and $V_{sn}$ are the intra- and interband matrix elements of the scattering potential, respectively. The latter is caused by the 
short-range scatterers with the momentum transfer $\sim \hbar/a_0 \gg p_{\rm F}$, where $a_0$ has an atomic scale, therefore the average product is assumed to be independent of $\varphi_{\bm p}$ and $\varphi_{\bm p'}$~\cite{ST_PRB_2011}. We emphasize that the obtained correction, Eq.~\eqref{dW}, is responsible for the effect of $E_z$-electric field on the 
scattering by pyramid-like scatters discussed in the model of the photon drag effect, see Sec.~\ref{section_modeldrag}.

Using the derived scattering probability $W_{\bm p' \bm p}$ we solve the Boltzmann equation~\eqref{Boltzmann} and obtain the $\bm r$-independent correction to the distribution function, $\delta f_{\bm p}$, which allows us to calculate the photon drag  current given by Eq.~\eqref{j}. With this we will search for the correction to the distribution function being responsible for the \textit{dynamical} alignment momentum, $f^{(da)}_{\bm p}$. First we find the linear in $\bm E_\parallel$ solution given by
\begin{equation} 
\label{fE}
	f^{(E)}_{\bm p}(\bm r)=  -{df_0\over d\varepsilon_{\bm p}} {e \tau_{\rm tr} \over 1 - {\rm i}\omega\tau_{\rm tr}}   (\bm E_\parallel \cdot \bm v_{\bm p}),
\end{equation}
where $f_0$ is the Fermi-Dirac distribution function,  and the transport relaxation time $\tau_{\rm tr}$, determining the mobility of 2D Dirac fermions is related to the symmetric part of the scattering probability as ${\tau_{\rm tr}^{-1}=\sum_{\bm p'}W^{(0)}_{\bm p' \bm p}(1-\cos{\theta_{\bm p'\bm p}})}$. The photon wavevector is accounted by the space derivatives in the kinetic equation~\cite{GlazovGanichev_review}
\begin{equation}
\label{da}
{\bm v}_{\bm p} \cdot {\partial f^{(E)}_{\bm p}\over \partial \bm r} = 
{\rm i} ({\bm v}_{\bm p} \cdot {\bm q}) f^{(E)}_{\bm p} = \left( {\rm i}\omega - {1\over \tau_2} \right) f^{(da)}_{\bm p} .
\end{equation}
The time $\tau_2 $ being of the order of $\tau_{\rm tr}$ describes relaxation of the above discussed alignment of charge carrier momenta. It is defined as follows
\begin{equation}
	\tau_2^{-1}=\sum_{\bm p'}W^{(0)}_{\bm p'\bm p}(1-\cos{2\theta_{\bm p'\bm p}}).
\end{equation}
From Eq.~\eqref{da} we find the correction to the distribution function describing the dynamical alignment of momenta in the form
\begin{equation}
	f^{(da)}_{\bm p} = - {{\rm i} \tau_2 ({\bm v}_{\bm p} \cdot {\bm q}) \over 1 - {\rm i}\omega\tau_2} f^{(E)}_{\bm p}.
\end{equation}

Now we take into account the anisotropic scattering which manifests itself as a correction given by Eq.~\eqref{dW}. The $\bm r$-independent correction to the distribution function $\delta f_{\bm p}$ is found from the following equation~\cite{order}
\begin{equation}
	\sum_{\bm p'} \delta W_{\bm p' \bm p} \left(f^{(da)}_{\bm p} + f^{(da)}_{\bm p'}\right) = - {\delta f_{\bm p}\over \tau_{\rm tr}}.	
\end{equation}

Finally, from Eq.~\eqref{j} we find the trigonal photon drag current. Experiments reveal that the photocurrent in all samples is caused by the linearly polarized radiation. For excitation by oblique incidence in the $(xz)$ plane it is described by
\begin{align}
	\label{PDqx_final}
&j_x = T_\parallel q_x E_x E_z = {e\beta\lambda^w p_{\rm F}^2 \omega\tau_2 (\tau_{\rm tr}+\tau_2) \over 4(1+\omega^2\tau_2^2)} \sigma(\omega) q_x E_x E_z, \nonumber \\
&j_y = -T_\parallel q_x E_y E_z. 
\end{align}
Here, the high-frequency conductivity is given by the Drude expression for degenerate 2D carriers
\[
\sigma(\omega) = {e^2 \varepsilon_{\rm F}\tau_{\rm tr} \over 4\pi\hbar^2(1+\omega^2\tau_{\rm tr}^2)}.\]
We introduce the anisotropic scattering constant which is nonzero due to C$_{3v}$ symmetry of the studied system
\begin{equation}
\label{beta}
\beta = { \sum\limits_{n} \left< {\rm Im} \left(V_{sn} {z_{ns} V_{ss}^*} \right) \right>  /\varepsilon_n
	\over \left< |V(\bm p' - \bm p)|^2 \sin^2{\theta_{\bm p' \bm p}} \right>},
\end{equation}
where brackets in the denominator mean averaging over both scatterer positions and the scattering angle $\theta_{\bm p' \bm p}$.
The derived Eqs.~\eqref{PDqx_final} show that, in line with experiments, see Figs.~\ref{theta}, and the discussion in Sec.~\ref{section_discussion} the current is even in the angle of incidence and vanishes for normal incidence at which $q_x =0$.  For elastic scattering by Coulomb impurities, the relaxation times are related as $\tau_2 = \tau_{\rm tr}/3$. According to Ref.~\onlinecite{Olbrich2014}, in this case the photogalvanic current is given by
\begin{align}
\label{PGE_final}
	&j_x =\chi (E_x^2-E_y^2) = ev_0 {2\tau_{\rm tr}\over E_{\rm F}} \Xi \, \sigma(\omega)(E_x^2-E_y^2) ,\\
	&j_y=-2\chi E_x E_y.\nonumber
\end{align}
where $\Xi$ is the factor describing the C$_{3v}$ symmetry of the studied system. In the studied samples we can estimate $\Xi \sim 10^{-4} \ldots  10^{-5}$, see also Ref.~\onlinecite{Olbrich2014}.

\begin{figure}
	\includegraphics[width=0.9\linewidth]{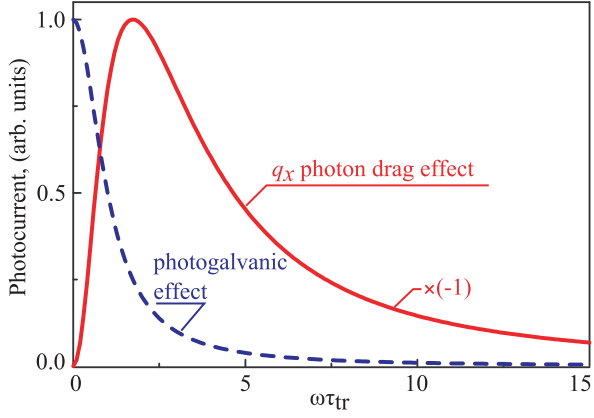}
	\caption{Frequency dependencies for photogalvanic and photon drag photocurrents calculated after Eq.~\eqref{PGE_final}, dashed curve and Eq.~\eqref{PDqx_final}, solid curve, respectively. The curves are normalized by the corresponding photocurrent maximum. The photon drag current is multiplied by (-1) because these terms have in experiments always the opposite sign.
	}
\label{freq_depend}
\end{figure}

The frequency dependencies of trigonal photon drag and photogalvanic currents are plotted in Fig.~\ref{freq_depend}. While photogalvanic current drops monotonously with frequency increase, the photon drag current has a maximum at $\omega\tau_{\rm tr} \approx 2$. At high frequencies, both linear photon drag and linear photogalvanic currents decrease as $\omega^{-2}$. The difference between the frequency dependencies may be the cause for the observed variation of the ratio between photon drag and photogalvanic currents varying in the range from about -2 to -15 for different samples and frequencies.

The ratio of the photon drag and photogalvanic currents can be estimated as 
\begin{equation}
	T_\parallel q/\chi \sim (\lambda^w p_{\rm F}^2/v_0)(\beta q \varepsilon_{\rm F}/\Xi).
\end{equation}
The first factor is a dimensionless degree of warping which can be of order of unity in our samples according to ARPES measurements,  Fig~\ref{arpes}. In our experiments, the trigonal photon drag current at oblique incidence is larger than the photogalvanic one. This allows us to estimate the interband scattering parameter $\beta$ defined by Eq.~\eqref{beta}: in the studied samples $\beta >10$~\AA/eV.
From the difference in sign between the photogalvanic and the trigonal photon drag currents systematically observed in experiment we conclude that the product $\beta\lambda^w$, describing the trigonal photon drag and the photogalvanic constant $\Xi$, are of opposite sign.

Finally we note, that the solution of the Boltzmann equation also yields the circular photon drag current proportional to the radiation helicity
\begin{align}
&j_x^{circ} =   {3e\beta\lambda^w p_{\rm F}^2 \tau_{\rm tr}(2-\omega^2\tau_{\rm tr}^2) \over 4+\omega^2\tau_{\rm tr}^2} \sigma(\omega) q_x {\rm i}(E_x E_z^*-E_x^* E_z), \nonumber \\
&j_y^{circ} = -{3e\beta\lambda^w p_{\rm F}^2 \tau_{\rm tr}(2-\omega^2\tau_{\rm tr}^2) \over 4+\omega^2\tau_{\rm tr}^2} \sigma(\omega) q_x {\rm i}(E_y E_z^*-E_y^* E_z). 
\end{align}
It follows from these expressions that the circular photon drag current is zero at $\omega\tau_{\rm tr}=\sqrt{2}$, a value at which the linear photon drag current is close to its maximum, Fig.~\ref{freq_depend}.  The vanishing contribution of the circular photocurrent in the vicinity of the $\omega\tau_{\rm tr} \approx 1$ -- the condition corresponding to our experiments -- may explain the fact  that in the studied frequency range no helicity dependent current has been detected.

\subsection{Microscopic theory of the photon drag effect due to $q_z$-component of the photon wavevector}

At last but not least we obtain the trigonal photon drag current caused by $q_z$-component of the photon wavevector. It is given by
\begin{align}
\label{Tz_def}
&	j_x= T_z q_z(|E_x|^2-|E_y|^2) , \\
&	j_y= -T_z q_z(E_xE_y^* + E_yE_x^*), \nonumber
\end{align}
where $T_z$ is a real constant. The equations reveal that the photocurrent can be excited by linearly polarized radiation and is sensitive to the polarization plane position in respect to the crystallographic axes.

The microscopic picture of the photocurrent generation can most conveniently be described in terms of the radiation magnetic field $\bm B$ rather than the transfer of the normal component of the photon wavevector $q_z$ to free carriers. Indeed, the latter is not possible in strictly 2D systems. Using the relation $\bm B = (c/\omega) \bm q\times \bm E$, we can rewrite Eq.~\eqref{Tz_def} as ${j_x \propto E_xB_y^* + E_yB_x^* + c.c.}$ and ${j_y \propto E_xB_x^* - E_yB_y^* + c.c.}$

In order to develop a microscopic theory for the photocurrent given by Eq.~\eqref{Tz_def}, we take into account the Lorentz force of the radiation magnetic field acting on the 2D carriers. The corresponding Hamiltonian has the following form:
\begin{equation}
	H_{\bm B}=  {ez\over m_0 c} (B_x p_y - B_y p_x).
\end{equation}
Taking into account both $H_{\bm B}$ and the warping perturbation Eq.~\eqref{warping} in the first order, we derive  the linear in $\bm B$ correction to the elastic scattering probability $\delta W_{\bm p' \bm p}^{(\bm B)}$.  It is obtained from $\delta W_{\bm p' \bm p}$, Eq.~\eqref{dW}, by substitution 
\[E_z \to {1 \over m_0c}[B_y(p_x+p_x') - B_x(p_y+p_y')] .
\]

Now we solve the Boltzmann kinetic Eq.~\eqref{Boltzmann} accounting the correction $\delta W_{\bm p' \bm p}^{(\bm B)}$ to the scattering probability. It has the form 
\begin{equation}
\label{f_EB}
	\sum_{\bm p'} \delta W_{\bm p' \bm p}^{(\bm B)} \left(f^{(E)}_{\bm p}-f^{(E)}_{\bm p'}\right) =- {\delta f^{(\bm B)}_{\bm p} \over \tau_{\rm tr}},
\end{equation}
where $\delta f^{(\bm B)}_{\bm p}$ is the correction to the distribution function linear in $\bm B$, and   the linear in $\bm E_\parallel$ correction $f^{(E)}_{\bm p}(\bm r)$ is given by Eq.~\eqref{fE}.

The linear in both $\bm E_\parallel$ and $\bm B_\parallel$ current density is calculated by Eq.~\eqref{j} with the above found correction $\delta f^{(\bm B)}_{\bm p}$.  The result has the form 
\begin{align}
\label{PDqz_final}
	& j_x = T_z q_z (E_x^2-E_y^2) = -{e\beta\lambda p_{\rm F}^2 \varepsilon_{\rm F} \over 4\omega m_0v_0^2} \sigma(\omega) q_z (E_x^2-E_y^2) , \nonumber \\
	& j_y = -2 T_z q_z E_x E_y.
\end{align}

Comparing the photocurrent amplitudes at normal and at oblique incidence, Eqs.~\eqref{PDqx_final} and~\eqref{PDqz_final}, we obtain 
\begin{equation}
T_z/T_\parallel \sim {1 + (\omega\tau_{\rm tr})^2 \over (\omega\tau_{\rm tr})^2} {\varepsilon_{\rm F} \over m_0v_0^2}.
\end{equation}
This estimate demonstrates that, because the radiation magnetic field  affects elastic scattering weaker  than the electric field, $T_z$ is substantially smaller than $T_\parallel$. Since $m_0v_0^2 \sim 10$~eV, the factor  $\varepsilon_{\rm F} / m_0v_0^2$ is in the order of $10^{-1}$ to $10^{-2}$ for our samples. This estimation explains why in all our experiments  $T_\parallel \gg T_z$. The smallest value obtained are 40 times for sample BST127 and 80 for sample BST641. In both cases the samples were excited by radiation of $f=3.3$~THz. For other conditions the ratio was even larger or the $q_z$-related photon drag contribution was not detectable. The only reason why we were able to detect such a small contribution at all, is that photon drag due to \textit{in-plane} wavevector vanishes at normal incidence, whereas that caused by $q_z$ component achieves its maximum. Finally we note that as $T_z$ constant has the frequency dependence different from $T_\parallel$. In particular, the role of $q_z$-photon drag current is enhanced at small frequencies ${\omega\tau_{\rm tr} \ll 1}$.

\section{Conclusion}
To summarize, our experiments on a large set of \textit{n}- and \textit{p}-type (Bi$_{1-x}$Sb$_{x}$)$_{2}$Te$_{3}$ three-dimensional topological insulators  demonstrated that normal incident THz radiation results in the photogalvanic current induced in the surface states. At oblique incidence, however,  in particular at large angles of incidence, it is outweighed by the photon drag effect. The developed microscopic model and theory show that the photon drag photocurrent is caused by the \textit{dynamical} momentum alignment by time and space dependent radiation electric field and implies the difference in the scattering probabilities for different half periods of the electromagnetic wave. Both photocurrents observed even at room temperature, stem from scattering events and, therefore, can be applied to study the high frequency conductivity in TI.

\appendix
\section{}
\label{appendix1}

Phenomenological analysis for C$_{3v}$ symmetry at linear polarization of radiation accounting for the photogalvanic and photon drag effects yields  even in $\theta$ photocurrents~\eqref{yz1} which can be conveniently written in the following form:
\begin{equation}
\label{A1}
j_x + {\rm i}j_y =  (\chi + T_z q_z) (E_x-{\rm i}E_y)^2 +  T_\parallel (q_x-{\rm i}q_y)(E_x-{\rm i}E_y)E_z.
\end{equation}

We consider oblique incidence with an incidence plane at an angle $\psi$ with the (${yz}$) plane, and  $\alpha$ is an angle between the radiation electric vector and the incidence plane (${\alpha=0}$ corresponds to $p$ polarization).
In these notations we have:
\begin{equation}
q_x-{\rm i}q_y = -{\rm i}q \sin{\theta} {\rm e}^{-{\rm i}\psi},
\quad
q_z = -q \cos{\theta},
\end{equation}
\begin{equation}
E_x-{\rm i}E_y = -(t_s \sin{\alpha} + {\rm i} t_p \cos{\alpha} \cos{\theta}) E_0 {\rm e}^{-{\rm i}\psi},
\end{equation}
\begin{equation}
E_z = t_pE_0 \sin{\theta}\sin{\alpha}.
\end{equation}
Here  $E_0$ is the electric field amplitude in vacuum, and  $t_s$, $t_p$ are Fresnel transmission coefficients for $s$ and $p$
polarizations. 

Substitution of the wavevector and electric field components into Eq.~\eqref{A1} yields the photocurrent in the following form:
\begin{equation}
j_x = j^{\rm off} -\cos{\left(2\alpha-\gamma \right)}\sqrt{(A_c\cos{2\psi})^2 + (A_s\sin{2\psi})^2}E_0^2,
\end{equation}
\begin{equation}
j_y = \sin{\left(2\alpha-\gamma'\right)\sqrt{(A_s\cos{2\psi})^2 + (A_c\sin{2\psi})^2}}E_0^2.
\end{equation}
Here the amplitudes are functions of the incidence angle:
\begin{equation}
A_c={1\over 2} [(\chi-T_z q \cos{\theta}) (t_s^2+t_p^2\cos^2{\theta})+ T_\parallel q t_p^2\sin^2{\theta} \cos{\theta}],
\end{equation}
\begin{equation}
A_s = t_st_p\cos{\theta}(\chi-T_z q \cos{\theta}) + T_\parallel q t_st_p\sin^2{\theta}/2,
\end{equation}
the phase shifts are given by:
\begin{equation}
\tan{\gamma} = {A_s \over A_c}\tan{2\psi},
\qquad
\tan{\gamma'} = {A_c \over A_s}\tan{2\psi},
\end{equation}
and the $\alpha$-independent offset photocurrent is:
\begin{equation}
j^{\rm off} = -{\sin^2{\theta}\over 2} \left[(\chi-T_z q \cos{\theta}) f(\theta) + T_\parallel q t_p^2\cos{\theta} \right]E_0^2,
\end{equation}
where
\begin{align}
&	f(\theta) \equiv {t_p^2\cos^2{\theta}-t_s^2 \over \sin^2{\theta}} \\
&	= {t_s t_p (t_p\cos{\theta}+t_s)\over 2\cos{\theta} } \left[ 1+ {n\cos{\theta}-1 \over n^2 \left(1+\sqrt{1-\sin^2{\theta}/n^2}\right)}\right] . \nonumber
\end{align}

Taking into account that in our structures ${T_\parallel q \gg \chi, T_zq}$ we have at small $\theta$:
\begin{align}
j_x = & -\cos{(2\alpha - 2\psi)} (\chi-T_zq + T_\parallel q \theta^2/2) t^2 E_0^2 + j^{\rm off},\\
j_y = & \sin{(2\alpha - 2\psi)} (\chi-T_zq+ T_\parallel q \theta^2/2) t^2 E_0^2.
\end{align}
Here $t$ is the amplitude transmission coefficient for normal incidence, and 
\begin{equation}
j^{\rm off} = - T_\parallel q \theta^2 t^2 E_0^2/2.
\end{equation}

\acknowledgments We thank M.\,M.~Glazov  and S.\,A.~Tarasenko, for fruitful discussions. The support from the DFG priority program SPP1666 and Virtual Institute for Topological Insulators, the Elite Network of Bavaria (K-NW-2013-247), and the Russian Foundation of Basic Research is gratefully acknowledged.


\begin{thebibliography}{99}



\bibitem{HasanKane2010} M. Z. Hasan and C. L. Kane, 
Rev. Mod. Phys. \textbf{82}, 3045 (2010).

\bibitem{Moore2010} J. E. Moore, 
Nature \textbf{464}, 194 (2010).

\bibitem{QiZhang2011}  X. L. Qi and S. C. Zhang,
Rev. Mod. Phys. \textbf{83}, 1057 (2011).

\bibitem{newreview} J. H. Bardarson and J. E. Moore, 
Rep. Prog. Phys. \textbf{76}, 056501 (2013). 

\bibitem{book2013} B. A. Bernevig, 
\textit{Topological Insulators and Superconductors}
(University Press Group Ltd, 2013)

\bibitem{book2013b} 
\textit{Topological Insulators Contemporary Concepts of Condensed Matter Science}
eds. M. Franz, and L. Molenkamp, 
(Elsevier Ltd, Oxford, 2013)

\bibitem{book2013c} S.-Q. Shen, 
\textit{Topological Insulators: Dirac Equation in Condensed Matters}
(Springer Series in Solid-State Sciences, 2013)

\bibitem{book2015} F. Ortmann, S. Roche, S. O. Valenzuela, Foreword by L. W. Molenkamp, 
\textit{Topological Insulators: Fundamentals and Perspectives}
(Wiley-VCH Verlag GmbH \& Co 2015)

\bibitem{book2015b}
\textit{Topological Insulators: The Physics of Spin Helicity in Quantum Transport} 
ed. Gregory Tkachov,  
(Pan Stanford Publishing Pte Ltd, 2015)


\bibitem{Ganichev2003} S. D. Ganichev and W. Prettl, 
topical review, J. Phys.: Condens. Matter, \textbf{15}, R935 (2003).

\bibitem{Ivchenko2008} E. L. Ivchenko and S. D. Ganichev, 
\textit{Spin Photogalvanics in Spin Physics in Semiconductors,} 
ed. M. I. Dyakonov 
(Springer 2008). 

\bibitem{book_Ivchenko} E. L. Ivchenko, 
\textit{Optical Spectroscopy of Semiconductor Nanostructures} 
(Alpha Science Int., Harrow, UK, 2005).


\bibitem{GlazovGanichev_review} M. M. Glazov and S. D. Ganichev,
Phys. Reports \textbf{535},  101 (2014).

helicity controlled photocurrent
\bibitem{Hosur2011} P. Hosur, 
Phys. Rev. B \textbf{83}, 035309 (2011).

\bibitem{McIver2012_2} J. W. McIver, D. Hsieh, H. Steinberg, P. Jarillo-Herrero, and N. Gedik, 
Nature Nanotechn. \textbf{7}, 96 (2012).

\bibitem{Junck2013} A. Junck, G. Refael, and F. von Oppen,
Phys. Rev. B \textbf{88}, 075144 (2013).

\bibitem{Duan2014} J. Duan, N. Tang, X. He, Y. Yan, S. Zhang, X. Qin, X. Wang, X. Yang, F. Xu, Y. Chen, W. Ge, and B. Shen, 
Sci. Rep. \textbf{4}, 4889 (2014).

\bibitem{Shikin2015} A.M. Shikin, A.A. Rybkina, I.I. Klimovskikh, M.V. Filianina, K.A. Kokh,
O.E. Tereshchenko, P.N. Skirdkov, K.A. Zvezdin, and A.K. Zvezdin
arXiv:1511.05663v1 (2015).

\bibitem{Olbrich2014} P. Olbrich, L.E. Golub, T. Herrmann, S.N. Danilov, H. Plank, V.V. Bel'kov, G. Mussler, Ch. Weyrich, C.M. Schneider, J. Kampmeier, D. Gr\"utzmacher, L. Plucinski, M. Eschbach, and S.D. Ganichev, 
Phys. Rev. Lett. \textbf{113}, 096601 (2014).

\bibitem{Braun2015} L. Braun, G. Mussler, A. Hruban, M. Konczykowski, M. Wolf, T. Schumann, M. M\"unzenberg, L. Perfetti, T. Kampfrath, 
arXiv:1511.00482 (2015).

\bibitem{Zhu2015} L.-G. Zhu, B. Kubera, K. F. Mak, and J. Shan, 
Sci. Rep. \textbf{5}, 10308 (2015).

\bibitem{Kastl2012} Ch. Kastl, T. Guan, X. Y. He, K. H. Wu, Y. Q. Li, and A. W. Holleitner, 
Appl. Phys. Lett. \textbf{101}, 251110 (2012).

\bibitem{Kastl2015_1} Ch. Kastl, Ch. Karnetzky, H. Karl, and A. W. Holleitner, 
Nature Comm. \textbf{6}, 6617 (2015).

\bibitem{Kvon2014} Z. D. Kvon, K.-M. Dantscher, C. Zoth, D. A. Kozlov, N. N. Mikhailov, S. A. Dvoretsky, and S. D. Ganichev, 
JETP Lett. \textbf{99}, 290 (2014).

\bibitem{Kaladzhyan2015} V. Kaladzhyan, P. P. Aseev, and S. N. Artemenko, 
Phys. Rev. B \textbf{92}, 155424 (2015).

\bibitem{Bas2012} D. A. Bas, K. Vargas-Velez, S. Babakiray, T. A. Johnson, P. Borisov, T. D. Stanescu, D. Lederman, and A. D. Bristow, 
Appl. Phys. Lett. \textbf{106}, 041109 (2015).

\bibitem{Muniz2014} R. A. Muniz and J. E. Sipe, 
Phys. Rev. B \textbf{89}, 205113 (2014).

\bibitem{Tanaka2014} Y. Onishi, Z. Ren, M. Novak, K. Segawa, Y. Ando, and K. Tanaka, 
arXiv:1403.2492 (2014).

\bibitem{Lee2016} H. C. Lee, 
Physica E
 \textbf{79}, 44
 (2016)

\bibitem{McIver2012} J. W. McIver, D. Hsieh, S. G. Drapcho, D. H. Torchinsky, D. R. Gardner, Y. S. Lee, and N. Gedik, 
Phys. Rev. B \textbf{86}, 035327 (2012).

\bibitem{Kitagawa2011} T. Kitagawa, T. Oka, A. Brataas, L. Fu, and E. Demler, 
Phys. Rev. B \textbf{84}, 235108 (2011).

\bibitem{Dora2012} B. Dora, J. Cayssol, F. Simon, and R. Moessner, 
Phys. Rev. Lett. \textbf{108}, 056602 (2012).

\bibitem{Olbrich2013} P. Olbrich, C. Zoth, P. Vierling, K.-M. Dantscher, G.V. Budkin, S.A. Tarasenko, V.V. Bel'kov, D.A. Kozlov, Z.D. Kvon, N.N. Mikhailov, S.A. Dvoretsky, and S.D. Ganichev, 
Phys. Rev. B \textbf{87}, 235439 (2013).

\bibitem{Dantscher2015} K.-M. Dantscher, D.A. Kozlov, P. Olbrich, C. Zoth, P. Faltermeier, M. Lindner, G. V. Budkin, S. A. Tarasenko, V. V. Bel'kov, Z.D. Kvon, N. N. Mikhailov, S.A. Dvoretsky, D. Weiss, B. Jenichen, and S. D. Ganichev, 
Phys. Rev. B \textbf{92}, 165314 (2015).

\bibitem{Zoth2014} C. Zoth, P. Olbrich, P. Vierling, K.-M. Dantscher, V.V. Bel'kov, M.A. Semina, M.M. Glazov, L.E. Golub, D.A. Kozlov, Z.D. Kvon, N.N. Mikhailov, S.A. Dvoretsky, and S.D. Ganichev,
Phys. Rev. B \textbf{90}, 205415 (2014).

\bibitem{Semenov2012} Y. G. Semenov, X. Li, and K. W. Kim, 
Phys. Rev. B \textbf{86}, 201401 (2012).

\bibitem{Li2014} X. Li, Yu. G. Semenov, and K. W. Kim, 
Appl. Phys. Lett. \textbf{104}, 061116 (2014).

\bibitem{Yao2012} J. D. Yao, J. M. Shao, S. W. Li, D. H. Bao, and G. W. Yang, 
Sci. Rep. \textbf{5}, 14184 (2015).

\bibitem{Egorova2015} S. G. Egorova, V. I. Chernichkin, L. I. Ryabova, E. P. Skipetrov, L. V. Yashina, S. N. Danilov, S. D. Ganichev, and D. R. Khokhlov, 
Scientific Reports \textbf{5}, 11540 (2015).


\bibitem{Checkelsky2009} J. G. Checkelsky, Y. S. Hor, M. H. Liu, D. X. Qu, R. J. Cava, and N. P. Ong,
Phys. Rev. Lett. \textbf{103}, 246601 (2009).

\bibitem{TaskinAndo2009} A. A. Taskin and Y. Ando,
Phys. Rev. B \textbf{80}, 085303 (2009).

\bibitem{Analytis2010_PRB}  J. G. Analytis, J. H. Chu, Y. Chen, F. Corredor, R. D. McDonald, Z. X. Shen, and I. R. Fisher,
Phys. Rev. B \textbf{81}, 205407 (2010).

\bibitem{Barreto2014} L. Barreto, L. K\"uhnemund, F. Edler, Ch. Tegenkamp, J. Mi, M. Bremholm, B. B. Iversen, Ch. Frydendahl, M. Bianchi, and P. Hofmann
Nano Lett. \textbf{14}, 7
 (2014).

\bibitem{Ren2010} Z. Ren, A. A. Taskin, S. Sasaki, K. Segawa, and Y. Ando
Phys. Rev. B \textbf{82}, 241306 (2010)

\bibitem{Qu2010} D.-X. Qu, Y. S. Hor, J. Xiong, R. J. Cava, N. P. Ong
Science \textbf{329}, 1189792 (2010)

\bibitem{ref-ternaries-Zhang} J. Zhang,	C.-Z. Chang, Z. Zhang, J. Wen, X. Feng, K. Li, M. Liu, K. He, L. Wang, W. Chen,	Q.-K. Xue, X. M. Wang,        
Nat. Commun. \textbf{2}, 574 (2011).

\bibitem{ref-ternaries} C. Weyrich, M. Dr\"ogeler, J. Kampmeier, M. Eschbach, G. Mussler, T. Merzenich, T. Stoica, I. E. Batov, J. Schubert, L. Plucinski, B. Beschoten, C. M. Schneider, C. Stampfer, D. Gr\"utzmacher, and Th. Sch\"apers, 
arxiv 1511.00965v2 (2015).

\bibitem{ref-pn} M. Eschbach, E. Młyńczak, J. Kellner, J. Kampmeier, M. Lanius, E. Neumann,	C. Weyrich,	M. Gehlmann, P. Gospodarič,	S. D\"oring, G. Mussler,	N. Demarina, M. Luysberg, G. Bihlmayer, Th. Sch\"apers, L. Plucinski, S. Bl\"ugel,	M. Morgenstern,	C. M. Schneider, and D. Gr\"utzmacher, 
Nat. Commun. \textbf{6}, 8816 (2015). 

\bibitem{vdW} Y. Liu, and M. Weinert, and L. Li,
Phys. Rev. Lett \textbf{108}, 115501 (2012).

\bibitem{Borisova2012} S. Borisova, J. Krumrain, M. Luysberg, G. Mussler, and D. Gr\"utzmacher,
Cryst. Growth Des. \textbf{12}, 6098
 (2012).

\bibitem{Borisova2013} S. Borisova, J. Kampmeier, M. Luysberg, G. Mussler and D. Gr\"utzmacher,
Appl. Phys. Lett. \textbf{103}, 081902 (2013).

\bibitem{Plucinski2013} L. Plucinski, A. Herdt, S. Fahrendorf, G. Bihlmayer, G. Mussler, S. D\"oring, J. Kampmeier, F. Matthes, D. E. Bürgler, D. Gr\"utzmacher, S. Bl\"ugel, and C. M. Schneider, 
Appl. Phys. \textbf{113}, 053706 (2013).

\bibitem{ARPES_2} Y. Xia, D. Qian, D. Hsieh, L. Wray, A. Pal, H. Lin, A. Bansil, D. Grauer, Y. S. Hor, R. J. Cava, and M. Z. Hasan,
Nature Physics \textbf{5}, 398 (2009).

\bibitem{ARPES_3}  D. Hsieh, Y. Xia, D. Qian, L. Wray, J. H. Dil, F. Meier, J. Osterwalder, L. Patthey, J. G. Checkelsky, N. P. Ong, A. V. Fedorov, H. Lin, A. Bansil, D. Grauer, Y. S. Hor, R. J. Cava, and M. Z. Hasan,
Nature (London) \textbf{460}, 1101 (2009). 

\bibitem{Kampmeier2015} J. Kampmeier, S. Borisova, L. Plucinski, M. Luysberg, G. Mussler, and D. Gr\"utzmacher, 
Cryst. Growth Des., \textbf{15}, 1 (2015).

\bibitem{book} S. D. Ganichev and  W. Prettl, 
\textit{Intense Terahertz Excitation of Semiconductors}
(Oxford Univ. Press, Oxford, 2006).

\bibitem{Tunnel1993} S.~D.~Ganichev, W.~Prettl, and P.~G.~Huggard,
Phys. Rev. Lett.  {\bf 71}, 3882 (1993).

\bibitem{DX1995} S.~D.~Ganichev, I.~N.~Yassievich, W.~Prettl, J.~Diener,  B.~K.~Meyer and K.~W.~Benz,
Phys. Rev. Lett. {\bf  75},  1590 (1995).

\bibitem{Tunnelingfrequency1998} S.~D.~Ganichev, E.~Ziemann, Th.~Gleim, W.~Prettl, I.~N.~Yassievich, V.~I.~Perel, I.~Wilke, and E.~E.~Haller,
Phys. Rev. Lett. {\bf 80}, 2409 (1998).

\bibitem{Ganichev1999} S. D. Ganichev
Physica B \textbf{273-274}, 737 (1999).


\bibitem{Ziemann2000}  E. Ziemann, S. D. Ganichev, I. N. Yassievich, V. I. Perel, and W. Prettl,
J. Appl. Phys. {\bf 87}, 3843 (2000).


\bibitem{Ganichev84p20} S. D. Ganichev, Y. V. Terent'ev, and I. D. Yaroshetskii, 
Pisma Zh. Tekh. Fiz. \textbf{11}, 46 (1985) [Sov. Tech. Phys. Lett. \textbf{11}, 20 (1989)].


\bibitem{BelkovSSTlateral} V. V. Bel'kov,  and S. D. Ganichev,
Semicond. Sci. Technol. \textbf{23}, 114003 (2008).

\bibitem{footnote1} The polarization independent offsets $C$ and $C'$, being in most samples much smaller than $A(f)$, may be caused by the non perfectly flat surface, which locally reduces the symmetry of the surface states and allows a polarization independent contribution to the both effects~\cite{Olbrich2014}.

\bibitem{footnote2} The difference in the photocurrent magnitudes $A(f)$ for front and back illuminations may be additionally affected by the unequal photocurrent contributions excited in the top and  interface surfaces separated by the bulk material, e.g., due to different scattering times.	

\bibitem{Resonantinversion2003} S.D.~Ganichev, V.~V.~Bel'kov, Petra~Schneider, E.~L.~Ivchenko,
S.A.~Tarasenko,  W.~Wegscheider, D.~Weiss,    D.~Schuh, E.V.~Beregulin  and  W.~Prettl, 
Phys.~Rev.~B {\bf 68}, 035319 (2003).

\bibitem{SGEopt2003} S.D.~Ganichev, Petra~Schneider, V.~V.~Bel'kov, E.~L.~Ivchenko, S.A.~Tarasenko, W.~Wegscheider, D.~Weiss, D.~Schuh, B.~N.~Murdin, P.~J.~Phillips,  C.~R.~Pidgeon, D.~G.~Clarke, M.~Merrick,  P.~Murzyn,  E.~V.~Beregulin, and W.~Prettl,
Phys.~Rev.~B. R {\bf 68}, 081302 (2003).

\bibitem{Diehlcircdrag2007} H.~Diehl, V.A. Shalygin, V.V.~Bel'kov,  Ch.~Hoffmann,  S.N.~Danilov, T.~Herrle, S.A.~Tarasenko,  D.~Schuh, Ch.~Gerl, W.~Wegscheider, W.~Prettl, and S.D.~Ganichev,
New J. Physics \textbf{9},  
349 (2007).

\bibitem{Karch2010} J. Karch, P. Olbrich, M. Schmalzbauer, C. Zoth, C. Brinsteiner, M. Fehrenbacher,  U. Wurstbauer, M. M. Glazov, S. A. Tarasenko, E. L. Ivchenko, D. Weiss, J. Eroms, R. Yakimova, S. Lara-Avila, S. Kubatkin, and S. D. Ganichev,
Phys. Rev. Lett. \textbf{105}, 227402 (2010).


\bibitem{3authors} S.D.~Ganichev, E.~L.~Ivchenko, and W.~Prettl,
Physica E {\bf 14}, 166 (2002).

\bibitem{GaN_LPGE} W. Weber, L. E. Golub, S. N. Danilov, J. Karch, C. Reitmaier, B. Wittmann, V. V. Bel'kov, E. L. Ivchenko, Z. D. Kvon, N. Q. Vinh, A. F. G. van der Meer, B. Murdin, and S. D. Ganichev, 
Phys. Rev. B \textbf{77}, 245304 (2008).

\bibitem{NatPhys09} H. Zhang, C.-X. Liu, X.-L. Qi, X. Dai, Z. Fang, and S.-C. Zhang,
Nature Phys. \textbf{5}, 438 (2009).

\bibitem{alignment} The stationary correction to the distribution function $f_{\bm p}(t)$ is obtained by writing it as an expansion in powers of the electric field $f_{\bm p}(t) = f_0 + f_{\bm p}^{(1)}(t) +  f_{\bm p}^{(2)}$ with the oscillating in time term $f_{\bm p}^{(1)}(t) \propto \exp{(-{\rm i}\omega t)}$ and the stationary term  $f_{\bm p}^{(2)} \propto |\bm E|^2$ being \textit{second order}  in the electric field. 

\bibitem{Belinicher-Strurman-UFN} V. I. Belinicher and B. I. Sturman,
Sov. Phys. Usp. \textbf{23}, 199 (1980).

\bibitem{alignment2} The oscillating in time term $f_{\bm p}^{(1)}(t)$, see \cite{alignment}, does not disturb the balance of scattered carriers, and consequently, does not results in a $dc$ current.

\bibitem{perelpinskii73}
V.~I. Perel' and Ya.~M. Pinskii, 
\newblock {Sov. Phys. Solid State}, {\bf 15}, 688 (1973). 

\bibitem{hex_warping_H} L. Fu,
Phys. Rev. Lett. \textbf{103}, 266801 (2009).

\bibitem{lambda2} Note that, in spite of the fact that $H_w \propto \lambda^w$, the correction to the electron energy is quadratic in $\lambda^w$ because linear in $\lambda^w$ terms are forbidden by time-inversion symmetry, so that $\delta\varepsilon_{\bm p} = - [(\lambda^w)^2 p^5/4v_0] \cos 6 \varphi_{\bm p}$. 

\bibitem{backscattering} S. Das Sarma, S. Adam, E. H. Hwang, and E. Rossi,
Rev. Mod. Phys. \textbf{83}, 407
 (2011).

\bibitem{ST_PRB_2011} S. A. Tarasenko, 
Phys. Rev. B \textbf{83}, 035313 (2011).

\bibitem{order} Note that if one first take into account $\delta W_{\bm p' \bm p}$ then the correction to the distribution function contains $E_zE_x^*$-like bilinear combinations of the electric field amplitudes which are $\bm r$-independent, and, hence, its gradient is zero.


%


\end{thebibliography}
\end{document}